\documentclass{article}

\usepackage[svgnames, x11names]{xcolor}

\newcommand{\ysnoted}[1]{}

\usepackage[normalem]{ulem} 

\usepackage{listings}

\lstdefinelanguage{XML}
{
	basicstyle=\ttfamily\footnotesize,
	morestring=[b]",
	moredelim=[s][\bfseries\color{Maroon}]{<}{\ },
	moredelim=[s][\bfseries\color{Maroon}]{</}{>},
	moredelim=[l][\bfseries\color{Maroon}]{/>},
	moredelim=[l][\bfseries\color{Maroon}]{>},
	morecomment=[s]{<?}{?>},
	morecomment=[s]{<!--}{-->},
	commentstyle=\color{gray},
	stringstyle=\color{blue},
	identifierstyle=\color{red}
}
\usepackage[pdftex]{graphicx}
\DeclareGraphicsExtensions{.pdf}

\usepackage[cmex10]{amsmath}
\usepackage{amssymb}
\usepackage{mathtools}

\usepackage[]{subfig}

\usepackage{algorithmicx}
\usepackage{algpseudocode}
\usepackage[ruled]{algorithm}
\definecolor{light-gray}{gray}{0.75}
\algrenewcommand{\algorithmiccomment}[1]{\hskip3em{{\footnotesize \textcolor{light-gray}{$\blacktriangleright$}}} #1}

\usepackage{multirow}

\usepackage{hyperref}

\usepackage{xspace}
\usepackage[nocompress]{cite}

\usepackage{enumitem}

\usepackage[english]{babel}
\usepackage{blindtext}

\begin{document}
	%
	\title{RIoTBench: A Real-time IoT Benchmark for Distributed Stream Processing Platforms}
	%
	%
	%
	%
	
	\author{Anshu Shukla, Shilpa Chaturvedi and Yogesh Simmhan \\
		\normalsize{\emph{Department of Computational and Data Sciences}}\\
		\normalsize{\emph{Indian Institute of Science (IISc), Bangalore 560012, India}}\\
		\normalsize{\emph{Email: shukla@grads.cds.iisc.ac.in, shilpa@grads.cds.iisc.ac.in, simmhan@cds.iisc.ac.in}}}
	
	\date{}
	\maketitle
	
	\begin{abstract}
		The Internet of Things (IoT) is an emerging technology paradigm where millions of sensors and actuators help monitor and manage, physical, environmental and human systems in real-time. The inherent closed-loop responsiveness and decision making of IoT applications make them ideal candidates for using low latency and scalable stream processing platforms. Distributed Stream Processing Systems (DSPS) hosted on Cloud data-centers are becoming the vital engine for real-time data processing and analytics in any IoT software architecture. But the efficacy and performance of contemporary DSPS have not been rigorously studied for IoT applications and data streams. Here, we develop \emph{RIoTBench}, a Real-time IoT Benchmark suite, along with performance metrics, to evaluate DSPS for streaming IoT applications. The benchmark includes $27$ common IoT tasks classified across various functional categories and implemented as reusable micro-benchmarks. Further, we propose four IoT application benchmarks composed from these tasks, and that leverage various dataflow semantics of DSPS. The applications are based on common IoT patterns for data pre-processing, statistical summarization and predictive analytics. These are coupled with four stream workloads sourced from real IoT observations on smart cities and fitness, with peak streams rates that range from $500-10,000~messages/sec$ and diverse frequency distributions. We validate the RIoTBench suite for the popular Apache Storm DSPS on the Microsoft Azure public Cloud, and present empirical observations. This suite can be used by DSPS researchers for performance analysis and resource scheduling, and by IoT practitioners to evaluate DSPS platforms.
	\end{abstract}

\section{Introduction}

\label{sec:intro}
Internet of Things (IoT) is a technology paradigm wherein ubiquitous sensors numbering in the billions will able to \emph{monitor} physical infrastructure and environment, human beings and virtual entities in real-time, \emph{process} both real-time and historic observations, and \emph{take actions} that improve the efficiency and reliability of systems, or the comfort and lifestyle of society. The technology building blocks for IoT have been ramping up over for a decade, with research into pervasive and ubiquitous computing~\cite{zaslavsky2013internet}, and sensor networks~\cite{chandrasekaran:cidr:2003} forming precursors. Recent growth in the capabilities of high-speed mobile (e.g., 3G/4G) and \emph{ad hoc} (e.g., Bluetooth) networks~\cite{ericsson-report}, smart phones and devices, affordable sensing and crowd-sourced data collection~\cite{data:city}, Cloud data-centers, and Big Data analytics platforms have all converged to the current inflection point for IoT.  

Existing IoT deployments in vertical domains such as \emph{Smart Power Grids}~\cite{simmhan:cise:2013} and \emph{health and fitness monitoring}~\cite{fitness-iot} already have millions of sensing and actuation points that constantly stream observations and trigger responses. 
The IoT stack for such domains is tightly integrated to serve specific needs, but typically operates on a closed-loop \emph{Observe Orient Decide Act (OODA)} cycle~\cite{perera:ieee:2014}, where sensors communicate time-series observations of the (physical or human) system to the Cloud for analysis, and the resulting analytics drives recommendations that are enacted on the system to improve it, which is again observed and so on. In fact, this \emph{closed-loop responsiveness} is one of the essential and distinguishing design characteristics of IoT applications, compared to other Big Data domains.


This low-latency cycle makes it necessary to process data streaming from sensors at fine spatial and temporal scales, in \emph{real-time}, to derive actionable intelligence. In particular, this streaming analytics has be to done at massive scales (millions of sensors, thousands of events per second) from across distributed sensors, requiring large computational resources. \emph{Cloud computing} offers a natural platform for scalable processing of the observations at globally distributed data centers, and sending a feedback response to the IoT system at the edge. 

Recent \emph{Big Data platforms} like Apache Storm~\cite{toshniwal:sigmod:2014}, Spark Streaming~\cite{zaharia:usenix:2012} and Flink~\cite{carbone2015flink} provide an intuitive dataflow programming model for composing such streaming applications, with a scalable, low-latency execution engine designed for commodity clusters and Clouds. 
These \emph{Distributed Stream Processing Systems (DSPS)} are becoming essential components of any IoT stack to support online analytics and decision-making for IoT applications. DSPS provide the ability to compose a dataflow graph of user-defined tasks that can process a continuous stream of opaque messages on distributed resources. This flexibility allows DSPS to incorporate a wide variety of business logic for real-time processing and online analytics necessary for a diverse and emerging domain like IoT. In fact, reference IoT solutions from Cloud providers like Amazon AWS\footnote{https://aws.amazon.com/iot/how-it-works/} and Microsoft Azure\footnote{https://www.microsoft.com/en-in/server-cloud/internet-of-things/overview.aspx} include their proprietary stream and event processing engines as part of the IoT software architecture.

Shared-memory stream processing systems~\cite{chandrasekaran:cidr:2003, aurora} have been investigated over a decade back for wireless sensor networks, with benchmarks such as \emph{Linear Road}~\cite{arasu:vldb:2004} being proposed. But there has not been a detailed review of, or benchmarks for, \emph{distributed} stream processing for IoT. IoT encompasses multiple domains, and applications go well beyond traditional social network and web traffic workloads for which DSPS were designed for~\cite{toshniwal:sigmod:2014}. They include a swathe of generalizable tasks for data pre-processing, statistical summarization and predictive analytics, as well as analytics for specific IoT application areas like Smart Transportation or health. As such, the efficacy and performance of contemporary DSPS have not been rigorously studied for \emph{IoT applications and data streams}. One reason is the absence of a well-defined IoT benchmark that realistically captures the domain features, exercises the unique compositional capabilities of DSPS, and validates them on real data streams. We address this gap in this paper.

This paper extends our prior published work, significantly increasing both the breadth and depth of the benchmark suite~\cite{shukla:tpctc:2016}. We add 14 new tasks to the earlier 13 tasks, including in new categories; 
two new streaming dataflow applications, besides updating the earlier two as well; and two new data workloads from the smart grid and personal fitness domains. We also include support for spatial scaling to increase the number of sensor streams, in addition to the temporal scaling used earlier to increase the stream rates. These make our benchmark comprehensive.

Specifically, we make the following contributions in this article:
\begin{enumerate}
	\item We classify different \emph{characteristics} of streaming applications, their composition semantics, and their data sources, in \S~\ref{sec:features}. 
	\item Then, in \S~\ref{sec:features:iot}, we propose \emph{categories of tasks} that are essential for IoT applications and the \emph{key features of input data streams} they operate upon.
	\item We identify \emph{performance metrics} of DSPS that are necessary to meet the latency and scalability needs of streaming IoT applications, in \S~\ref{sec:metrics}.
	\item We propose the \textbf{RIoTBench} real-time IoT benchmark for DSPS based on representative \emph{micro-benchmark tasks}, drawn from the above categories, in \S~\ref{sec:benchmark}. We design four reference \emph{IoT applications} that span Data pre-processing, Statistical analytics and Predictive Analytics, and are composed from these tasks. We also identify \emph{four real-world streams} with different distributions as workloads on which to evaluate them.
	\item Lastly, we validate the proposed benchmark suite for the popular \emph{Apache Storm} DSPS, and present empirical results for the same in \S~\ref{sec:results}.
\end{enumerate}

Our contributions benefit two classes of audience. One, for \emph{developers and users in IoT domains}, RIoTBench offers a set of realistic IoT tasks and applications that they can customize and configure to help evaluate candidate DSPS platforms for their performance and scalability needs. Two, for \emph{researchers on Big Data}, it provides a reference micro and application benchmark, along with datasets, that can be used as a baseline to uniformly compare the impact of their research advances in resource management, scalability and resiliency for DSPS on the emerging IoT domain.

\section{Background and Related Work}
\label{sec:related}
Stream processing systems allow users to compose applications as a dataflow graph, with task vertices having some user-defined logic and streaming edges passing messages between the tasks. The systems then run the applications continuously over incoming data streams. 
%
Early Data Stream Management Systems (DSMS) extended Database Management Systems (DBMS) to support by sensor network applications, that have similarities to IoT~\cite{carney:vldb:2002,chen:sigmod:2000,babu:record:2001}. They supported continuous query languages with operators such as join and aggregation similar to SQL, but with a temporal dimension using time and tuple window operations. These have been extended to distributed implementations~\cite{balazinska:tods:2008,biem:sigmod:2010} and, more recently, complex event processing (CEP) engines for detecting sequences and patterns~\cite{cugola:csur:2012}.  

Contemporary Distributed Stream Processing Systems (DSPS) like Apache Storm, Spark Streaming, Flink and Yahoo S4~\cite{toshniwal:sigmod:2014,zaharia:usenix:2012,carbone2015flink,neumeyer:icdmw:2010} were designed using Big Data fundamentals -- running on commodity clusters and Clouds, offering weak scaling, ensuring robustness, and supporting fast data processing over thousands of events per second. Unlike DSMS, DSPS do not support native query operators and instead allow users to plug in their own logic composed as dataflows that are executed on a cluster. Event processing and querying can be higher-level abstractions on top of these~\footnote{Apache Trident, http://storm.apache.org/releases/1.0.1/Trident-tutorial.html}. While developed for web and social network applications, such fast data platforms have found use in financial markets, astronomy, and particle physics. IoT is one of the more recent domains to consider them. 

%


There are design and architectural differences even within DSPS, which we highlight as part of our characterization. The types of programming semantics supported can vary, and determines the flexibility in composition. Spark Streaming uses micro-batch processing in contrast to per-tuple processing in Storm, with consequences trade-offs between latency and throughput. As a result, it is important to qualitatively and quantitatively evaluate these frameworks for specific application domains, and the distributed platform they target. Understanding the common set of feature dimensions and performance metrics, in addition to the actual IoT benchmark definitions, is necessary for fair comparison across the DSPS. We discuss these later for DSPS operating on Clouds to support IoT applications.

\subsection{DSPS Benchmarks}
Work on DSMS spawned the \emph{Linear Road Benchmark (LRB)}~\cite{arasu:vldb:2004} that was proposed as an application benchmark. In the scenario, the DSMS had to evaluate toll and traffic queries over event streams from a virtual toll collection and traffic monitoring system. This has parallels with current smart transportation scenarios. 
However, there have been few studies or community efforts on benchmarking DSPS, other than individual evaluation of research prototypes against popular DSPS like Storm or Spark. 
These efforts define their own measures of success -- typically limited to throughput and latency -- and use generic workloads such as the Enron email dataset with empty operations (NoOps) as micro-benchmark to compare InfoSphere Streams ~\cite{nabi:streams:2014} and Storm. 



\emph{SparkBench}~\cite{agrawal:sparkbench:2015} is a framework-specific benchmark for Apache Spark, and includes four categories of applications from domains spanning Graph computation and SQL queries, with one on streaming applications supported by Spark Streaming. The benchmark metrics include CPU, memory, disk and network IO, with the goal of identifying tuning parameters to improve Spark's performance. 
%
\emph{CEPBen}~\cite{li:tpctc:2014} evaluates the performance of CEP systems based of the functional behavior of queries. It shows the degree of complexity of CEP operations like filter, transform and pattern detection. The evaluation metrics consider event processing latency, but ignore network overheads and CPU utilization. Further, CEP applications rely on a declarative query syntax to match event patterns rather than a dataflow composition based on user-logic provided by DSPS.

\emph{StreamBench}~\cite{lu:ucc:2014} is the closest work that partially addresses our goals. The authors propose 7 micro-benchmarks on 4 different synthetic workload suites generated from real-time web logs and network traffic to evaluate DSPS. Metrics including performance, durability and fault tolerance are proposed. The benchmark covers different dataflow composition patterns and common tasks like grep and wordcount, and compare Storm and Spark Streaming. 

The paper, while addressing the gap that existed in generalizable benchmarks DSPS, still falls short on several counts. 
It focuses on micro-benchmarks
and does not consider larger applications with more tasks and complex structures. 
Design patterns like 
duplicates and round-robin, and selectivity ratios 
are not explicitly considered. 
The benchmark does not cover a broad range of realistic input data rates either.
We address these gaps. At the same time, we do not emphasize durability or fault-tolerance metrics in our study, through these metrics can be added.


In contrast to these DSPS benchmarks, RIoTBench offers relevant micro- and application-level benchmarks for evaluating DSPS, specifically for \emph{IoT workloads} for which such platforms are increasingly being used. Our benchmark is designed to be \emph{platform-agnostic}, \emph{simple} to implement and execute within diverse DSPS, and \emph{representative} of both the application logic and the data stream workloads observed in IoT domains. This allows for the performance of DSPS to be independently and reproducibly verified for IoT applications. 




\subsection{Big Data and IoT Benchmarks}
There has been a slew of Big Data benchmarks that have been developed recently in the context of processing high volume (i.e., MapReduce-style) and enterprise/web data that complement our work. 
\emph{Hibench}~\cite{huang:hibench:2010} is a workload suite for evaluating Hadoop with popular micro-benchmarks like Sort, WordCount and TeraSort, MapReduce applications like Nutch Indexing and PageRank, and machine learning algorithms like K-means Clustering. \emph{BigDataBench}~\cite{gao:bigdatabench:2013} analyzes workloads from social network and search engines, and analytics algorithms like Support Vector Machine (SVM) over structured, semi-structured and unstructured web data. Both these benchmarks are general purpose workloads that do not target any specific  domain, but MapReduce platforms at large. 

\emph{BigBench}~\cite{ghazal:acm:2013} uses a synthetic data generator to simulate enterprise data found in online retail businesses. It combines structured data generation from the TPC-DS benchmark~\cite{nambiar:vldb:2006}, semi-structured data on user clicks, and unstructured data from online product reviews. Queries cover data \emph{velocity} by processing periodic refreshes that feed into the data store, \emph{variety} by including free-text user reviews, and \emph{volume} by querying over a large web log of clicks. We take a similar approach for benchmarking fast data platforms, targeting the IoT domain specifically and using real public data streams. 






\emph{Chronos}~\cite{gu:chronos:2015} is a recent work to generate and simulate streams for benchmarking. Their aim is to generate realistic input data streams with a distribution similar to given sample events. They use elastic infrastructure to generate events at high rates, and validate their work for telecom, advertising and stock market data. Their work is complementary to ours, as we propose dataflow patterns and applications, as well as representative datasets as part of benchmarks which are run at their native and scaled rates. Chronos can be used to stress these benchmarks further with larger inputs and faster rates.

There has been some recent work on benchmarking IoT applications. In particular, the generating large volumes of synthetic sensor data with realistic values is challenging, yet required for benchmarking. \emph{IoTAbench}~\cite{arlitt:icpe:2015} provides a scalable synthetic generator of time-series datasets. It uses a Markov chain model for scaling the time series with a limited number of inputs such that important statistical properties of the stream is retained in the generated data. They have demonstrated this for smart meter data. The benchmark also includes six SQL queries to evaluate the performance of different query platforms on the generated dataset. Their emphasis is on the data characteristics and content, which supplements our focus on evaluating the runtime aspects of the DSPS platform. 

\emph{CityBench}~\cite{ali:citybench:2015} is a benchmark to evaluate RDF stream processing systems. They include different generation patterns for smart city data, such as traffic vehicles, parking, weather, pollution, cultural and library events, with changing event rates and playback speeds. They propose fixed set of semantic queries over this dataset, with concurrent execution of queries and sensor streams. Here, the target platform is different (RDF database), but in a spirit as our work. 

Benchmarks for IoT hardware is also becoming important. \emph{IoT-Connect}~\cite{iotconnect} is an Industry-Standard Benchmark for Embedded Systems to analyze the behavior of micro-controllers with various connectivity interfaces 
like Bluetooth, Thread, LoRa, and WiFi. It also provides methods to determine the energy consumption for IoT devices. 






\section{Characteristics of DSPS Applications and Streams}
\label{sec:features}

In this section, we review the common application composition capabilities of DSPS, and the dimensions of the streaming applications that affect their performance on DSPS. These semantics help define and describe streaming IoT applications based on DSPS capabilities.



\subsection{Dataflow Composition Semantics}
\label{subsec:dataflow}
DSPS applications are composed as a \emph{dataflow graph}, where vertices are user provided \emph{tasks} and directed edges refer to \emph{streams of messages} that can pass between them. The graph need not be acyclic. Tasks in the dataflows can execute zero or more times, and a task execution usually depends on data-dependency semantics, i.e, when ``adequate'' inputs are available, the task executes. However, there are also more nuanced patterns that are supported by DSPS that we discuss.
\emph{Messages} (or events or tuples) from/to the stream are consumed/produced by the tasks. 
DSPS typically treat the messages as opaque content, and only the user logic may interpret the message content. However, DSPS may assign identifiers to messages for fault-tolerance and delivery guarantees, and some message attributes may be explicitly exposed as part of the application composition for the DSPS to route messages to downstream tasks. 

\emph{Selectivity ratio}, also called \emph{gain}, is the average number of output messages emitted by a task on consuming a unit input message, expressed as $\sigma= \text{\emph{input rate}}:\text{\emph{output rate}}$. Based on this, one can assess whether a task amplifies or attenuates the incoming message rate. It is important to consider this while designing benchmarks as it can have a multiplicative impact on downstream tasks. 




\begin{figure*}[t]
	\centering
	\includegraphics[width=\textwidth]{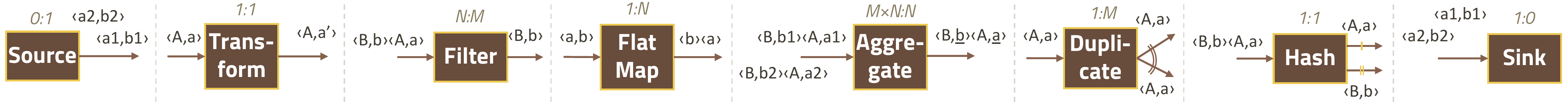}
	\caption{Common task patterns and semantics in streaming applications.}
	\label{fig:semantics}
\end{figure*}

There are message generation, consumption and routing semantics associated with tasks and their composition. Fig.~\ref{fig:semantics} captures the basic \emph{composition patterns} supported by modern DSPS. \texttt{Source} tasks have only outgoing edge(s), and these tasks encapsulate user logic to generate or receive the input messages that are passed to the dataflow. Likewise, \texttt{Sink} tasks have only incoming edge(s) and these react to the output messages from the application, say, by storing it or sending an external notification. 

\texttt{Transform} tasks, sometimes called \emph{Map}~\footnote{Spark Programming Guide, http://spark.apache.org/docs/latest/programming-guide.html}, generate one output message for every input message received ($\sigma=1:1$). Their user logic performs a transformation on the message, such as changing the units or projecting only a subset of attribute values. \texttt{Filter} tasks allow only a subset of messages that they receive to pass through, optionally performing a transformation on them ($\sigma=N:M$, $N \ge M$). Conversely, a \texttt{FlatMap} consumes one message and emits multiple messages ($\sigma=1:N$). An \texttt{Aggregate} pattern consumes a \emph{window} of messages, with the window width provided as a \emph{count} or a \emph{time} duration, and generates one or more messages that is an aggregation over each message window ($\sigma=N:1$). Specific DSPS may expose additional dataflow patterns as well.

When a task has multiple outgoing edges, routing semantics on the dataflow decide if an output message is \emph{duplicated} onto all the edges, or just one downstream task is selected for delivery, either based on a \emph{round robin} behavior or using a \emph{hash function} on an attribute in the outgoing message to decide the target task. 
Similarly, multiple incoming streams arriving at a task may be \emph{merged} into a single interleaved message stream for the task. Or alternatively, the messages coming on each incoming stream may be conjugated, based on order of arrival or an attribute exposed in each message, to form a \emph{joined} stream of messages. 
Other custom DSPS routing semantics may exist too.

There are additional dimensions of the streaming dataflow that can determine its performance on a DSPS. 
%
%
Tasks may be \emph{data parallel}, in which case, it can be allocated concurrent resources (threads, cores) to process messages in parallel by different instances the task. This is typically possible for tasks that do not maintain state across multiple messages. The \emph{number of tasks} in the dataflow graph indicates the size of the streaming application. Tasks are mapped to computing resources, and depending of their degree of parallelism and resource usage, it determines the cores/VMs required for executing the application. 
%
The \emph{length of the dataflow} is the latency of the critical (i.e., longest) path through the dataflow graph, if the graph does not have cycles. This gives an estimate of the expected latency for each message and also influences the number of network hops a message on the critical path has to take in the cluster. 

\subsection{Data Stream Characteristics}
We list a few characteristics of the input data streams that impact the runtime performance of streaming applications, and help classify IoT message streams. 

The \emph{input throughput} in messages/sec is the cumulative frequency at which messages enter the source tasks of the dataflow. Input throughputs can vary by application domain, and are determined both by the number of streams of messages and their individual rates. This combined with the dataflow selectivity will impact the load on the dataflow and its individual tasks, and determine the output throughput. 

\emph{Throughput distribution} captures the variation of input throughput over time. In real-world settings, the input data rate is usually not constant and DSPS need to adapt to this. There may be several common data rate distributions besides a \emph{uniform} one. There can be \emph{bursts} of data coming from a single sensor, or a coordinated set of sensors. A \emph{saw-tooth} behavior is seen in the ramp-up/-down before/after specific events. 
A \emph{Normal} distribution can occur for diurnal (day vs. night) stream sources, with \emph{bi-modal} variations capturing peaks during the morning and evening periods of human activity. 

Lastly, the \emph{message size} provides the average size of each message, in bytes. Often, the messages sizes remain constant for structured messages arriving from specific sensor or observation types, but may vary for free-text input streams or those that interleave messages of different types. This size help assess the communication cost of transferring messages in the dataflow.

\section{Characteristics of IoT Applications and Streams}
\label{sec:features:iot}
Here, we categorize IoT tasks, applications and data streams used within DSPS, based on the domain requirements. These, together with the patterns and semantics discussed in the previous section, offer a search space for defining dataflows and workloads that meaningfully and comprehensively validate IoT applications on DSPS. 

\subsection{Categories of IoT Tasks}
IoT covers a broad swathe of domains, many of which are rapidly developing. So, it is not possible to comprehensively capture all possible IoT application scenarios. However, DSPS have clear value in supporting the real-time processing, analytics, decision making and feedback that is intrinsic to most IoT domains. Here, we attempt to categorize these common processing and analytics tasks that are performed over real-time data streams. 

\textbf{Parse.} Messages are encoded on the wire in a standard text-based or binary representation by the stream sources, and need to be parsed upon arrival at the application. Text formats in particular require string parsing by the tasks, and are also larger in size on the wire. The tasks within the application may themselves retain the incoming format in their streams, or switch to another format or data model, say, by projecting a subset of the fields. They may also annotate and extend the number of fields. Industry-standard formats that are popular for IoT domains include CSV, XML, SenML and JSON text formats, and EXI and CBOR binary formats. For e.g., IETF's \emph{SenML (Sensor Markup Language)}~\cite{senml} can define an array of entries, where each entry is an object object that encapsulates attributes and their values, such as the unique identifier for the sensor, the time of measurement, and the current value, with the ability to model repetitions, relative time, etc. SenML serializations into JSON, XML and EXI are possible.

\textbf{Filter.} Messages may require to be filtered based on specific attribute values present in them, for data quality checks, to route a subset of message types to a part of the dataflow graph, or as part of their application logic. 
\emph{Value filters} such as min/max or band-pass filters check the numeric values of different observational fields from the sensors and can drop outliers.  Since IoT event rates may be high, more efficient \emph{Bloom filters} are a probabilistic structure that can be used to process large sets of discrete values with low space complexity but a small fraction of false positives. It can be used to detect invalid sensors or users in an incoming data streams. Filtering over text or media streams is also possible, but requires consideration like using text or video processing libraries. 

\textbf{Statistical Analytics.} Groups of messages within a sequential time or count window of a stream may require to be aggregated as part of the application. The aggregation function may be \emph{common mathematical operations} like average, count, minimum and maximum. They may also be \emph{higher order statistics} such finding outliers, quartiles, second and third order moments, and counts of distinct elements. Statistical \emph{data cleaning} like linear interpolation or denoising using Kalman filters are common for sensor-based data streams. Some tasks may maintain just local state for the window width (e.g., local average) while others may maintain state across windows (e.g., moving average). When the state size grows, here again approximate aggregation algorithms may be used. \emph{Distinct approximate count} is another such example of statistical tasks where we try to find approximate distinct values present in stream. 

\textbf{Predictive Analytics.} Predicting future behavior of the system based on past and current messages is an important part of IoT applications. Various statistical and machine-learning algorithms may be employed for predictive analytics over sensor streams. The \emph{predictions} may either use a recent window of messages to estimate the future values over a time or count horizon in future, or train models over streaming messages that are periodically used for predictions over the incoming messages.
Even simple techniques like \emph{interpolation} can be useful for replacing empty entries by interpolation over past values.  \emph{Classification} algorithms like decision trees, neural networks and na\"{i}ve Bayes can be trained to map discrete values to a category, which may lead to specific actions taken on the IoT system. External packages like Weka or R may be used by such tasks.
The \emph{training} itself can be an online task that is part of a DSPS dataflow. For e.g., ARIMA and linear regression use statistical methods to predict uni- or multi-variate attribute values, respectively. Also trained models can be updated on the fly within such forecasting tasks. 

\textbf{Pattern Detection.} Another class of tasks are those that identify patterns of behavior over several events. Unlike window aggregation which operate over static window sizes and perform a function over the values, pattern detection matches user-defined predicates on messages that may not be sequential or even span streams, and returns the matched messages. These are often modeled as \emph{state transition automata} or \emph{query graphs}. Common patterns include contiguous or non-contiguous sequence of messages with specific property on each message (e.g., high-low-high pattern over 3 messages), a join over two streams based on a common attribute value, or even semantic matching~\cite{zhou2016knowledge}. Complex Event Processing (CEP) engines like Siddhi~\cite{siddhi} may be embedded within the DSPS task to match such patterns.

\textbf{Visual Analytics.} Other than automated decision making, IoT applications often generate \emph{charts and animations} for consumption by end-users or system managers. These visual analytics may be performed at the client's browser using libraries like \texttt{D3.js}, in which case the processed data stream is aggregated and provided to the users. Alternatively, the streaming application may itself periodically generate such plots and visualizations as part of the dataflow, to be hosted on the web or pushed to the client. Charting and visualization libraries like \texttt{XChart}, \texttt{gnuplot} or \texttt{matplotlib} may be used for this purpose.  

\textbf{IO Operations.} Lastly, the IoT dataflow may need to access external storage or messaging services to access/push data into/out of the application. These may be to store or load trained models, archive incoming data streams, access historic data for aggregation and comparison, and subscribe to message streams or publish actions back to the system. These require access to \emph{file storage, SQL and NoSQL databases,} and \emph{publish-subscribe messaging systems}. Often, these may be hosted as part of the Cloud platforms themselves like Azure Storage. This also include writing files to local or remote disk, and optionally compressing or uncompressing them. Each of them have their own characteristics in term of latency, peak rate supported and resource usage.

\subsection{Categories of IoT Applications}

The tasks from the above categories, along with other domain-specific tasks, are composed together to form streaming IoT dataflow applications. These domain dataflows themselves fall into specific classes based on common use-case scenarios, and loosely map to the Observe-Orient-Decide-Act (OODA) phases.



\textbf{Extract-Transform-Load (ETL) and Archival} applications are front-line ``observation'' dataflows that receive and pre-process the data streams, and if necessary, archive a copy of the data offline. Pre-processing may perform data format transformations, normalize the units of observations, data quality checks to remove invalid data, interpolate missing data items, and temporally reorder messages arriving from different streams, annotate with the metadata. The pre-processed data may be archived to table storage, and passed onto subsequent dataflow for further analysis.

\textbf{Summarization and Visualization} applications perform statistical aggregation and analytics over the data streams to understand the behavior of the IoT system at a coarser granularity. Statistical analytics may include tasks such as finding approximate counts, identifying skewness in data distribution, and using linear regression for online trends. Such summarization can give the high-level pulse of the system, and help ``orient'' the decision making to the current situation. These tasks are often coupled with visualization tasks in the dataflow to present the summary status to end-users and decision makers.

\textbf{Prediction and Pattern Detection} applications use current information and historic models to help determine the future state of the IoT system, and ``decide'' if any reaction is required. They identify patterns of interest that may indicate the need for a correction, or forecasts based on current behavior that require preemptive actions. For e.g., a trend that indicates an unsustainably growing load on a smart power grid may cause a decision to preemptively shed load, or a detection that the heart-rate from a fitness watch is dangerously high may trigger a decision to reduce physical exertion. Model-based prediction applications are also coupled with batch or online dataflow applications that periodically re-train the models using observed data. 

\textbf{Classification and Notification} applications determine specific ``actions'' that are required and communicate them to the IoT system. Decisions may be mapped to specific actions, and the entities in the IoT system that can enact those are notified. These notifications can be delivered using SMS gateways, web service calls, or publish-subscribe brokers. For e.g., the need for load shedding in the power grid may map to notifying specific residents with a request for curtailment, or the need to reduce physical exertion may lead to a treadmill being notified to reduce the speed. The classification or case based reasoning systems may also require model training, like for predictive analytics.

\subsection{IoT Data Stream Characteristics}
IoT data streams are often generated by physical sensors that observe physical systems or the environment. As a result, they are typically time-series data that are generated periodically by the sensors. The sampling rate for these sensors may range from once a day to hundreds per second, depending on the domain. The number of sensors themselves may vary from a few hundred to millions as well. IoT applications like smart power grids can generate high frequency plug load data at thousands of messages/sec from a small cluster of residents, or low frequency data from a large set of sensors, such as in smart transportation or environmental sensing. As a result, we may encounter a wide range of input throughputs from $10^{-2}$ to $10^{5}$ messages/sec. In comparison, streaming web applications like Twitter deal with $6000$~tweets/sec from 300M users.

At the same time, this event rate itself may not be uniform across time. Sensors may also be configured to emit data only when there is a change in observed value, rather than unnecessarily transmitting data that has not changed. This helps conserve network bandwidth and power for constrained devices when the observations are slow changing. Further, if data freshness is not critical to the application, they may sample at high rate but transmit at low rates but in a burst mode. E.g. smart meters may collecting kWh data at 15~min intervals from millions of residents but report it to the utility only a few times a day, while the FitBit smart watch syncs with the Cloud every few minutes or hours even as data is recorded every few seconds.
Message variability also comes into play when human-related activity is being tracked. Diurnal or bimodal event rates are seen with single peaks in the afternoons, or dual peaks in the morning and evening. E.g., sensors at businesses may match the former while traffic flow sensors may match the latter.

There may also be a variety of observation types from the same sensor device, or different sensor devices generating messages. These may appear in the same message as different fields, or as different data streams. This will affect both the message rate and the message size. These sensors usually send well-formed messages rather than free-text messages, using standards like SenML. Hence their sizes are likely to be deterministic if the encoding format is not considered -- text formats tend to bloat the size and also introduce size variability when mapping numbers to strings. However, social media like tweets and crowd-sourced data are occasionally used by IoT applications, and these may have more variability in message sizes.

\section{Performance  Metrics}
\label{sec:metrics}

We identify and formalize commonly-used quantitative performance measures for evaluating DSPS for the IoT workloads. 

\textbf{Latency.} 
Latency for a message that is generated by a task is the time in seconds it took for that task to process one or more inputs to generate that message. If $\sigma=N:M$ is the selectivity for a task $T$, the time $\lambda^T_M$ it took to consume $N$ messages to \emph{causally produce} those $M$ output messages is the latency of the $M$ messages, with the \emph{average latency} per message given by $\overline{\lambda^T} = \frac{\lambda^T_M}{|M|}$. When we consider the average latency $\overline{\lambda}$ of the dataflow application, it is the average of the time difference between each message consumed at the source tasks and all its causally dependent messages generated at the sink tasks. 

The latency per message may vary depending on the input rate, resources allocated to the task, and the type of message being processed. 
While this task latency is the inverse of the mean throughput, the \emph{end-to-end latency} for the task within a dataflow will also include the network and queuing time to receive a tuple and transmit it downstream.

\textbf{Throughput.} 
The output throughput is the aggregated rate of output messages emitted out of the sink tasks, measured in messages per second. 
The throughput of a dataflow depends on the input throughput and the selectivity of the dataflow, provided the resource allocation and performance of the DSPS are adequate. Ideally, the output throughput $\omega^o = \sigma \times \omega^i$, where $\omega^i$ is the input throughput for a dataflow with selectivity $\sigma$. It is also useful to measure the \emph{peak throughput} that can be supported by a given application, which is the maximum stable rate that can be processed using a fixed quanta of resources. 

Both throughput and latency measurements are relevant only under \emph{stable conditions} when the DSPS can sustain a given input rate, i.e., when the latency per message and queue size on the input buffer remain constant and do not increase unsustainably.

\textbf{Jitter.} The ideal output throughput may deviate due to variable rate of the input streams, change in the paths taken by the input stream through the dataflow (e.g., at a \texttt{Hash} pattern), or performance variability of the DSPS. We use jitter to track the variation in the output throughput from the expected output throughput, defined for a time interval $t$ as, 
\[ J_t = \frac{\omega^o - \sigma \times \omega^i}{\sigma \times \overline{\omega^i}} \] where the numerator is the observed difference between the expected and actual output rate during interval $t$, and the denominator is the expected long term average output rate given a long-term average input rate $\overline{\omega^i}$. In the case of an ideal DSPS, jitter will tend toward zero, even if there are instantaneous changes in the input rate.

\textbf{CPU and Memory Utilization.} Streaming IoT dataflows are expected to be resource intensive, and the ability of the DSPS to use the distributed resources efficiently with minimal overhead is important. This also affects the VM resources and consequent price to be paid to run the application using the given stream processing platform. 
We track the CPU and memory utilization for the dataflow as the average of the CPU and memory utilization across all the VMs that are being used by the dataflow's tasks. The per-VM information can also help identify which VMs hosting which tasks are the potential bottlenecks and can benefit from data-parallel scale-out, and cases of over-allocation of resources. 

\section{\emph{RIoTBench} IoT Benchmark Suite}
\label{sec:benchmark}
We propose benchmark workloads to help evaluate the metrics discussed before for various DSPS. These benchmarks are in particular targeted for emerging IoT applications, to help them distinguish the capabilities of contemporary DSPS on Cloud computing infrastructure. The benchmarks themselves have two parts, the dataflow logic that is executed on the DSPS and the input data streams that they are executed for. We next discuss our choices for both.



\subsection{IoT Micro-benchmarks}

We propose a suite of common IoT tasks that span various categories we have identified and different streaming task patterns. These tasks form independent micro-benchmarks, and are further composed into application benchmarks later. The goal of the micro-benchmarks is to evaluate the performance of the DSPS for individual IoT tasks, and we measure the \emph{peak input throughput} that they can sustain on a unit computing resource as the performance metric. This offers a baseline for comparison with other DSPS, and can also inform resource scheduling decisions for more complex application dataflows composed using these tasks. 

\begin{table}[t]
	\footnotesize
	\centering
	\caption{IoT Micro-benchmark Tasks with different IoT Categories and DSPS Patterns.}
	\begin{tabular}{lccccc}
		\hline
		\textbf{Task Name} &~\textbf{Code}~&~\textbf{Category}~& \textbf{Pattern} &~\textbf{$\sigma$ Ratio}~&~\textbf{State}\\ \hline
		\hline
		
		Annotate & ANN & Parse &  Transform &1:1&No\\
		CsvToSenML & C2S & Parse &  Transform &1:1&No\\
		SenML Parsing~\cite{senml} & SML & Parse & Transform &1:1&No\\
		XML Parsing & XML & Parse & Transform &1:1&No\\
		\hline
		Bloom Filter~\cite{bloom:acm:1970} & BLF & Filter & Filter &1:0/1&No\\
		Range Filter & RGF & Filter & Filter &1:0/1 & No\\
		\hline
		Accumlator & ACC& Statistical  & Aggregate &N:1&Yes\\
		Average & AVG& Statistical  & Aggregate &N:1&Yes\\
		Distinct Appox. Count~\cite{durand:esa:2003} & DAC & Statistical & Transform &1:1&Yes\\
		Kalman Filter~\cite{kalman:asme:1959} & KAL & Statistical & Transform &1:1&Yes\\
		Second Order Moment~\cite{alon:stoc:1996} & SOM & Statistical & Transform &1:1&Yes\\
		\hline
		Decision Tree Classify~\cite{quinlan:ml:1986} & DTC  & Predictive & Transform &1:1&No\\
		Decision Tree Train & DTT  & Predictive & Aggregate  &N:1&No\\
		Interpolation & INP & Predictive & Transform &1:1&Yes\\
		Multi-var. Linear Reg. & MLR& Predictive & Transform &1:1&No\\
		Multi-var. Linear Reg. Train & MLT& Predictive & Aggregate &N:1&No\\
		Sliding Linear Regression & SLR& Predictive & Flat Map &N:M&Yes\\
		\hline
		Azure Blob D/L & ABD& IO & Source/Transform &1:1&No\\
		Azure Blob U/L & ABU& IO & Sink &1:1&No\\
		Azure Table Lookup & ATL& IO & Source/Transform &1:1&No\\
		Azure Table Range & ATR& IO & Source/Transform &1:1&No\\
		Azure Table Insert & ATI& IO & Transform &1:1&No\\
		MQTT Publish & MQP& IO & Sink &1:1&No\\
		MQTT Subscribe & MQS& IO & Sink &1:1&No\\
		Local Files Zip & LZP & IO & Sink &1:1&No\\
		Remote Files Zip & RZP & IO & Sink &1:1&No\\
		\hline
		MultiLine Plot~\cite{xchart:lib} & PLT& Visualization  & Transform &1:1&No\\
		\hline\hline
	\end{tabular}
	\label{tbl:tasts}
\end{table}


Table~\ref{tbl:tasts} lists the different micro-benchmark tasks, and their IoT categories, task patterns, and selectivity. These are grouped by their categories. The \emph{parse} category includes tasks that process standard text formats such as SenML and XML, and convert them to object formats, and also convert from a CSV format to a SenML form with additional semantics. The annotation task appends metadata content to an existing message based on an in-memory lookup for a unique ID present in the tuple. All these parse tasks transform messages from one form to another. 
The Bloom filter finds practical use in the \emph{filter} category for processing a large, discrete data space. It is trained with a white-list of valid sensor IDs that it will permit. The simple value range filter is used filtering in messages with observation fields that fall within a fixed upper and lower bound.

We have several tasks in the \emph{statistical analytics} category that perform aggregations and transformations. Basic statistics include a simple average of a single attribute's values over a count window, and a generic accumulator that buffers incoming messages based on a count window for use by other tasks. The second order moments over time-series values is another common statistics we implement. Estimating the frequencies a large range of streaming values can be memory intensive, and distinct approximate count performs a probabilistic count over the incoming messages while conserving memory. Lastly, the Kalman filter we provide is a popular denoising algorithm used for smoothing sensor data values in a time-series.

\emph{Predictive analytics} uses the Weka library to implement several common Machine Learning tasks. A multi-variate linear regression is included to predict one attribute's numerical value based on the values of one or more other in the message. This has both online training and online prediction tasks.  Similarly, the decision tree classifier is used for predict a class based on enumerated field values in the message, and also comes with a training and a classification task. Training for both these models happens over large, batched windows of messages. Interpolation and linear regression are standard techniques used for univariate time-series observation, and are available in the micro-benchmark suite. 

We have several \emph{IO} tasks for reading and writing to Microsoft Azure Cloud's file (blob) storage and NoSQL (table) storage. In addition, the common file operation of compressing a set of files is also available, with the source files being either on local disk or on the network. Publish/subscribe to/from an MQTT publish-subscribe broker for notifications and also included. Lastly, a single exemplar \emph{Visualization} task in the form of a Java XChart plotting library is present to accumulate and generate an image file.  

A micro-benchmark dataflow is composed for each of these tasks as a sequence of a source task, the benchmark task and a sink task. As can be seen, these tasks also capture different dataflow patterns such as transform, filter, aggregate, flat map, source and sink.

\subsection{IoT Application Benchmarks}
\begin{figure}[t]
	\centering
	\includegraphics[width=0.4\textwidth]{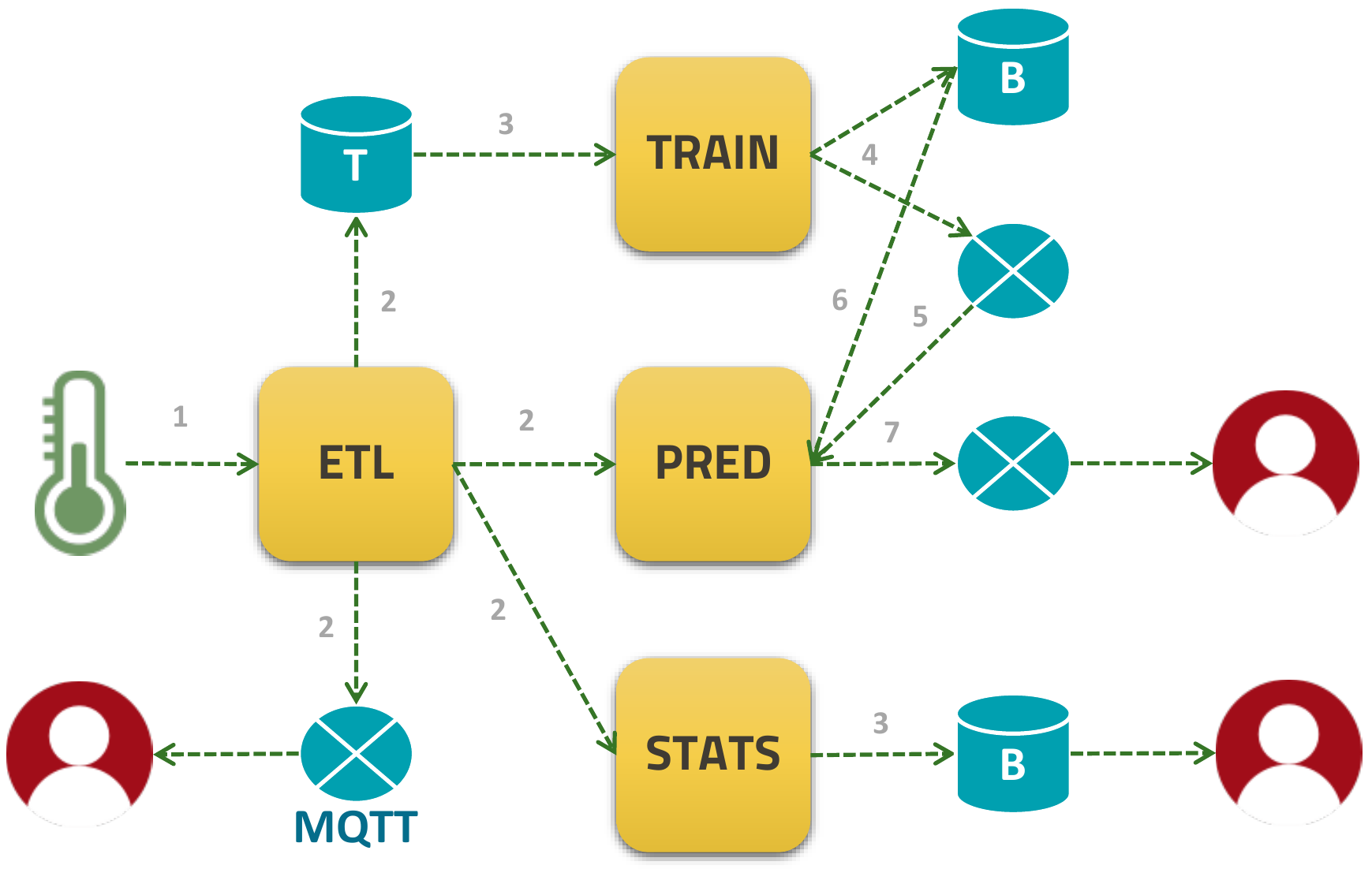}
	\caption{High-level logical interactions between different sensors, benchmark applications, platform services and users}
	\label{fig:app:flow}
\end{figure}

Application benchmarks are valuable in understanding how non-trivial and meaningful IoT applications behave on DSPS. Application dataflows for a domain are most representative when they are constructed based on real or realistic application logic, rather than synthetic tasks. In case applications use highly-custom logic or proprietary libraries, this may not be feasible or reusable as a community benchmark. However, many of the common IoT tasks we have defined earlier are naturally composable into application benchmarks that satisfy the requirements of a OODA decision making loop. 
%
%

Fig.~\ref{fig:app:flow} shows a high-level use case of such an IoT scenario that is generalizable to many domains such as smart power, transportation and fitness. This is achieved by the interaction between four different application dataflows we propose. Here, input streams from sensors in the domain arrive at a \emph{Extract-Transform-Load (ETL)} dataflow that performs data pre-processing and cleaning on the observations and archives it to Cloud table storage. Further, one output stream is published to the MQTT message broker so that clients interested in real-time monitoring can subscribe to it, while another copy is forked to the second dataflow which performs  \emph{Statistical Summarization (STATS)}. This application does higher order aggregation and plotting operations, and stores the generated plots to Cloud blob file storage, from where web-pages can load the visualization files on browsers. 

Concurrently, two dataflows support predictive analytics. \emph{Model Training (TRAIN)} periodically loads the archived data from the Cloud table store and trains forecasting models that are stored in the Cloud, and notifies the MQTT broker of an updated model being available. The \emph{Predictive Analytics (PRED)} dataflow subscribes to the broker and downloads the new models from the Cloud, and continuously operates over the pre-processed data stream from ETL to make predictions and classifications that can indicate actions to be taken on the domain. It then notifies the message broker of the predictions, that can independently be subscribed to by a user or device for action.

\begin{figure}[t]
	\centering
	\subfloat[Extraction, Transform and Load (ETL)]{%
		\includegraphics[width=0.7\textwidth]{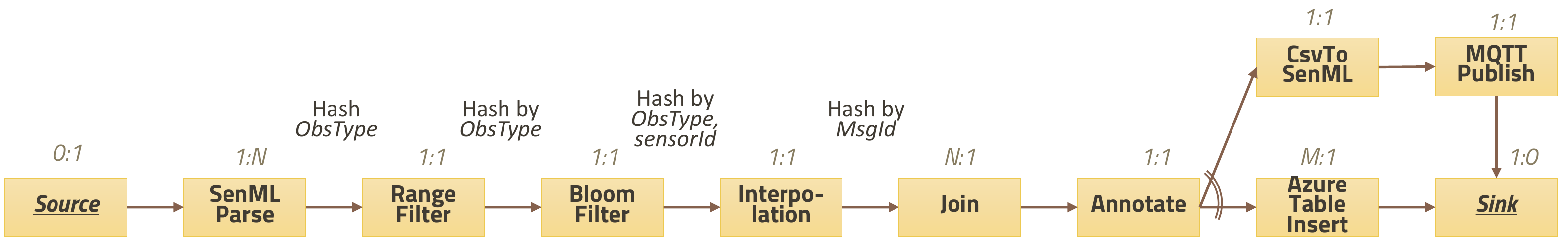}%
		\label{fig:app-etl}
	}\\
	\subfloat[Statistical Summarization (STATS)]{%
		\includegraphics[width=0.7\textwidth]{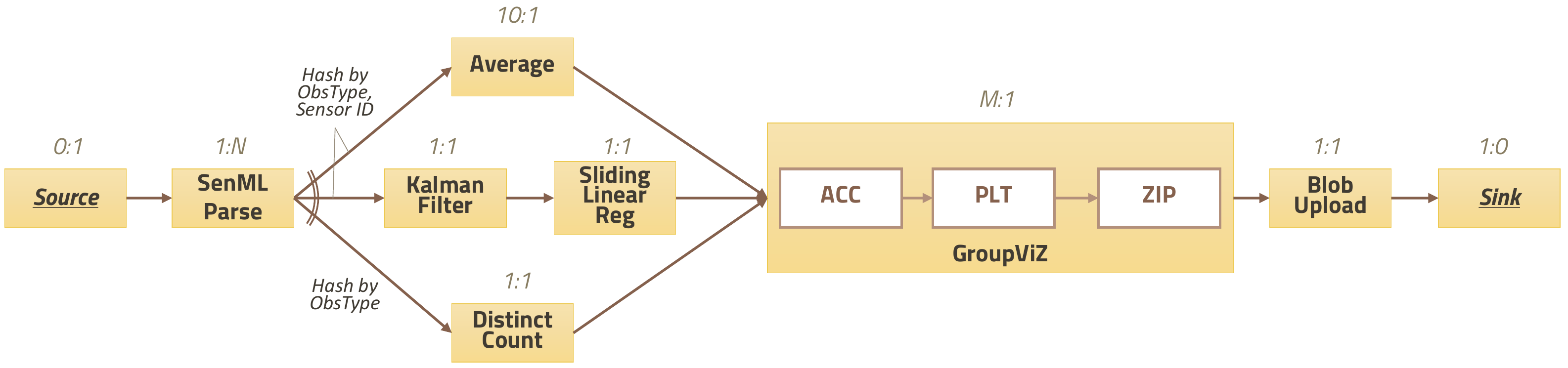}%
		\label{fig:app-stats}
	}\\
	\subfloat[Model Training (TRAIN)]{%
		\includegraphics[width=0.7\textwidth]{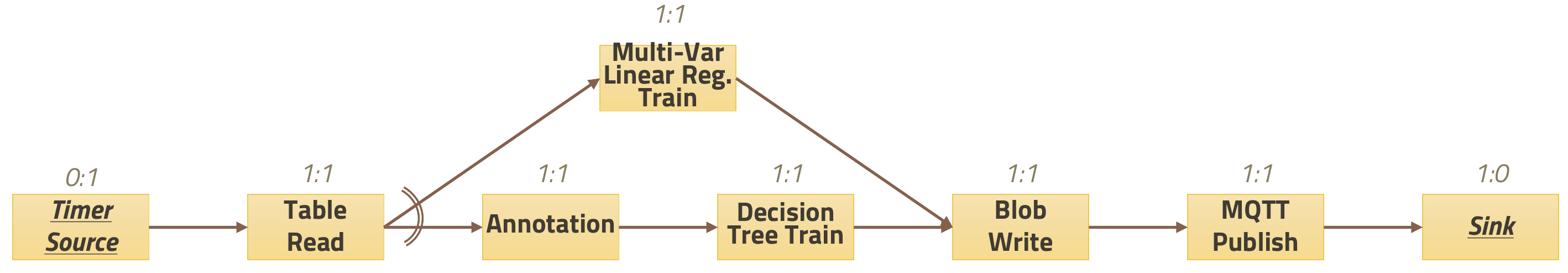}%
		\label{fig:app-train}
	}\\  
	\subfloat[Predictive Analytics (PRED)]{%
		\includegraphics[width=0.7\textwidth]{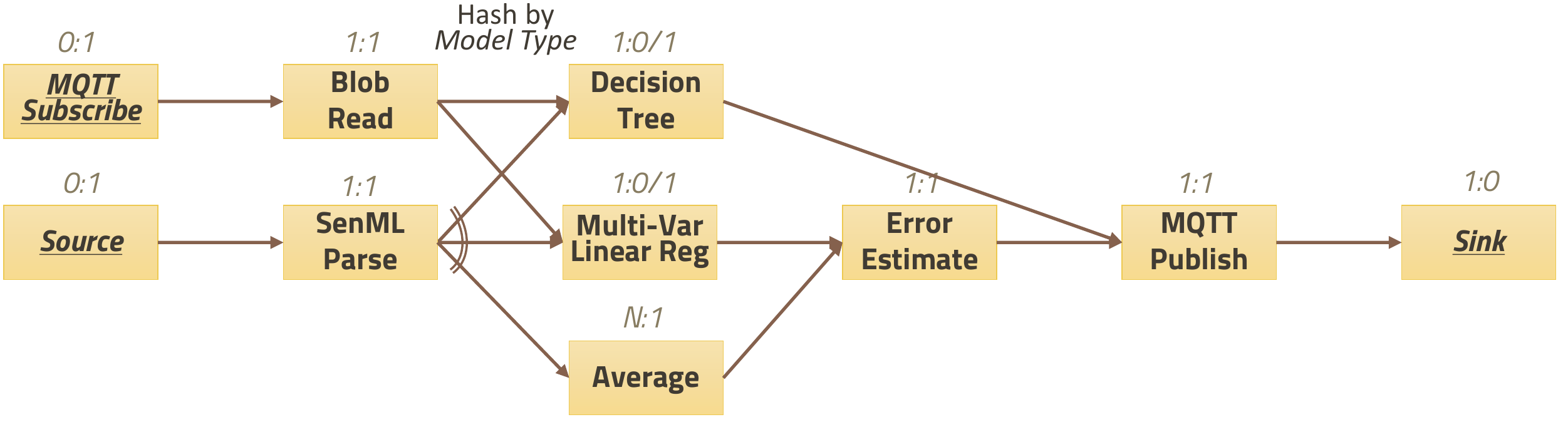}%
		\label{fig:app-pred}
	}
	\caption{Application benchmarks composed using the micro-benchmark tasks.}
	\label{fig:app}
\end{figure}

More specifically, \textbf{ETL} (Fig.~\ref{fig:app-etl}) ingests incoming data streams in SenML format, performs data filtering of outliers on individual observation types using a Range and Bloom filter, and subsequently interpolates missing values. It then annotates additional meta-data into the observed fields of the message and then inserts the resulting tuples into Azure table storage, while also converting the data back to SenML and publishing it to MQTT. 
A dummy sink task shown is used for logging purposes.

The \textbf{STATS} dataflow (Fig.~\ref{fig:app-stats}) parses the input messages that arrive in SenML format -- typically from the ETL, but kept separate here for modularity. It then performs three types of statistical analytics in parallel on individual observation fields present in the message: an average over a $10$ message window,  
Kalman filtering to smooth the observation fields followed by a sliding window linear regression, and an approximate count of distinct values that arrive. These three output streams are then grouped for each sensor IDs, plotted and the resulting image files zipped. These three tasks are tightly coupled and we combine them into a single meta-task for manageability, as is common. and the output file is written to Cloud storage for hosting on a portal.

The \textbf{TRAIN} (Fig.~\ref{fig:app-train}) application uses a timer to periodically (e.g., for every minute) trigger a model training run. Each run fetches data from the Azure table available since the last run and uses ti to train a Linear Regression model. In addition, these fetched tuples are also annotated to allow a Decision Tree classifier to be trained.  
Both these trained model files are then uploaded to Azure blob storage and their files URLs are published to the MQTT broker. 

The \textbf{PRED} (Fig.~\ref{fig:app-pred}) application subscribes to these notifications and fetches the new model files from the blob store, and updates the downstream prediction tasks. Meanwhile, the dataflow also consumes pre-processed messages streaming in, say from the ETL dataflow, and after parsing it 
forks it to the decision tree classifier and the multi-variate regression tasks. 
The classifier assigns messages into classes, such as good, average or poor, based on one or more of their field values, while linear regression predicts a numerical attribute value in the message using several others. The regression task also compares the predicted values against a moving average and estimates the residual error 
between them. 
The predicted classes, values and errors are published to the MQTT broker. 
The Appendix lists configuration parameters and attributes used for relevant tasks in the dataflows for different workloads we benchmark them on.

As such, these applications leverage many of the compositional capabilities of DSPS. The dataflows include \emph{single and dual sources}; tasks that are composed \emph{sequentially, task-parallel} and as \emph{combined} meta-tasks; \emph{stateful and stateless} tasks; and \emph{data parallel tasks} allowing for concurrent instances. Each message in the data streams contains multiple observation fields, but several of these tasks are applicable only on univariate streams and some are meaningful only from time-series data from individual sources. Thus, the initial parse task for ETL and STATS uses a \emph{flat map} pattern ($\sigma=1:N$, where N is number of observational fields) to create observation-specific streams early on. These streams are further passed to task instances, grouped by their observation type and optionally their sensor/meter ID using a \emph{hash pattern}.


\subsection{IoT Input Stream Workloads}
\begin{figure}[t]
	\centering
	\subfloat[CITY]{%
		\includegraphics[width=0.3\textwidth]{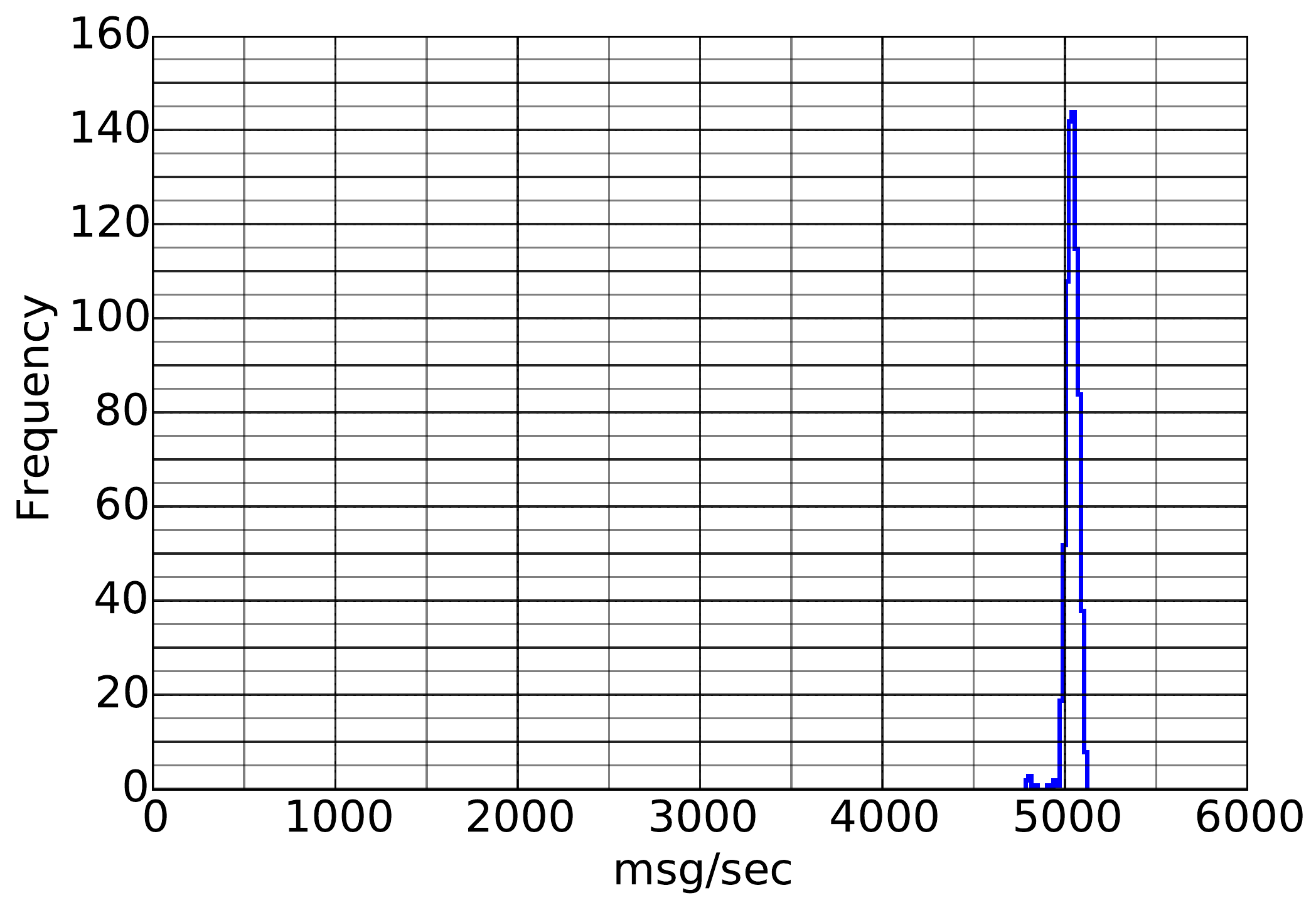}%
		\label{fig:data:sys}%
	}  
	\subfloat[TAXI]{%
		\includegraphics[width=0.3\textwidth]{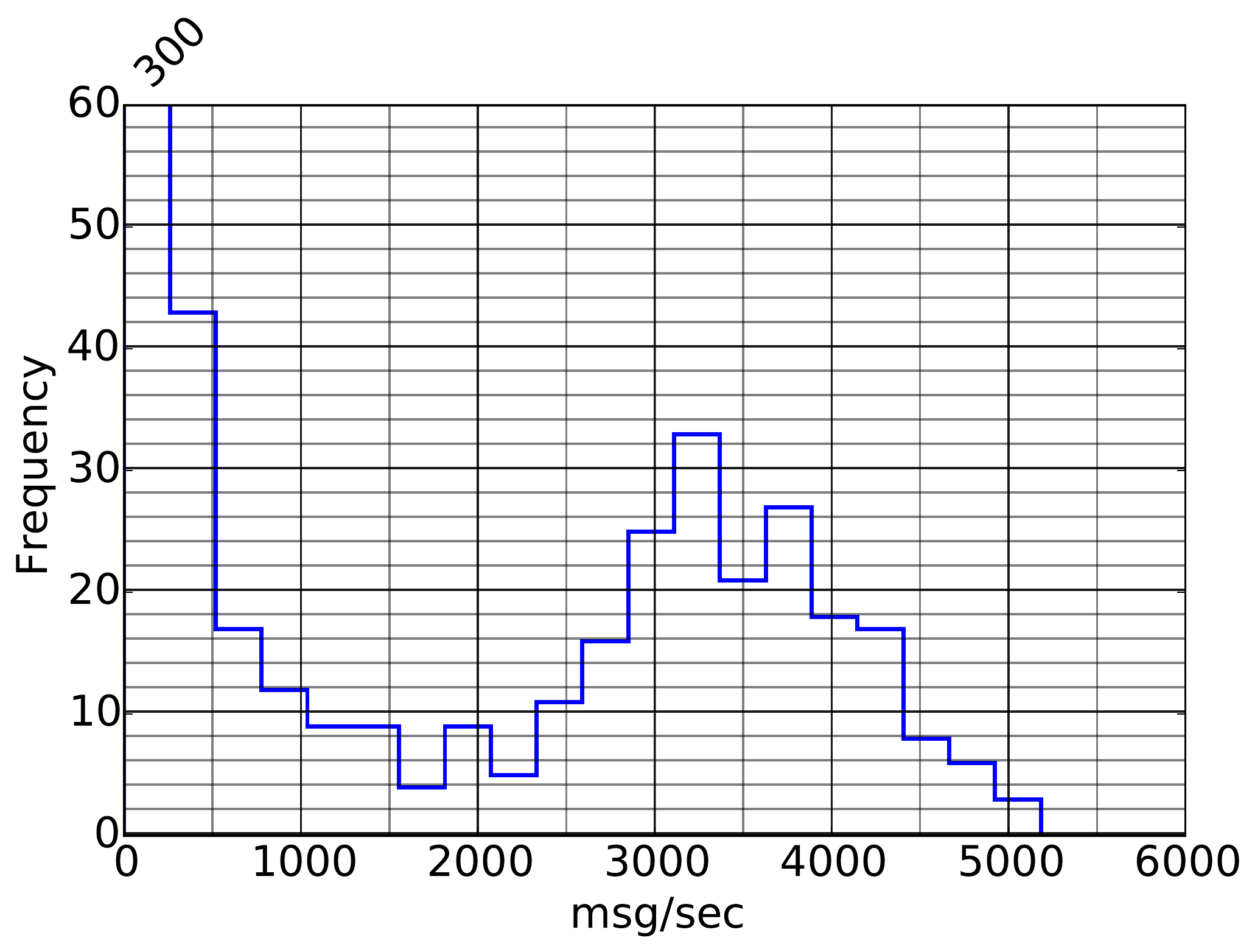}%
		\label{fig:data:taxi}%
	}\\  
	\subfloat[GRID]{%
		\includegraphics[width=0.3\textwidth]{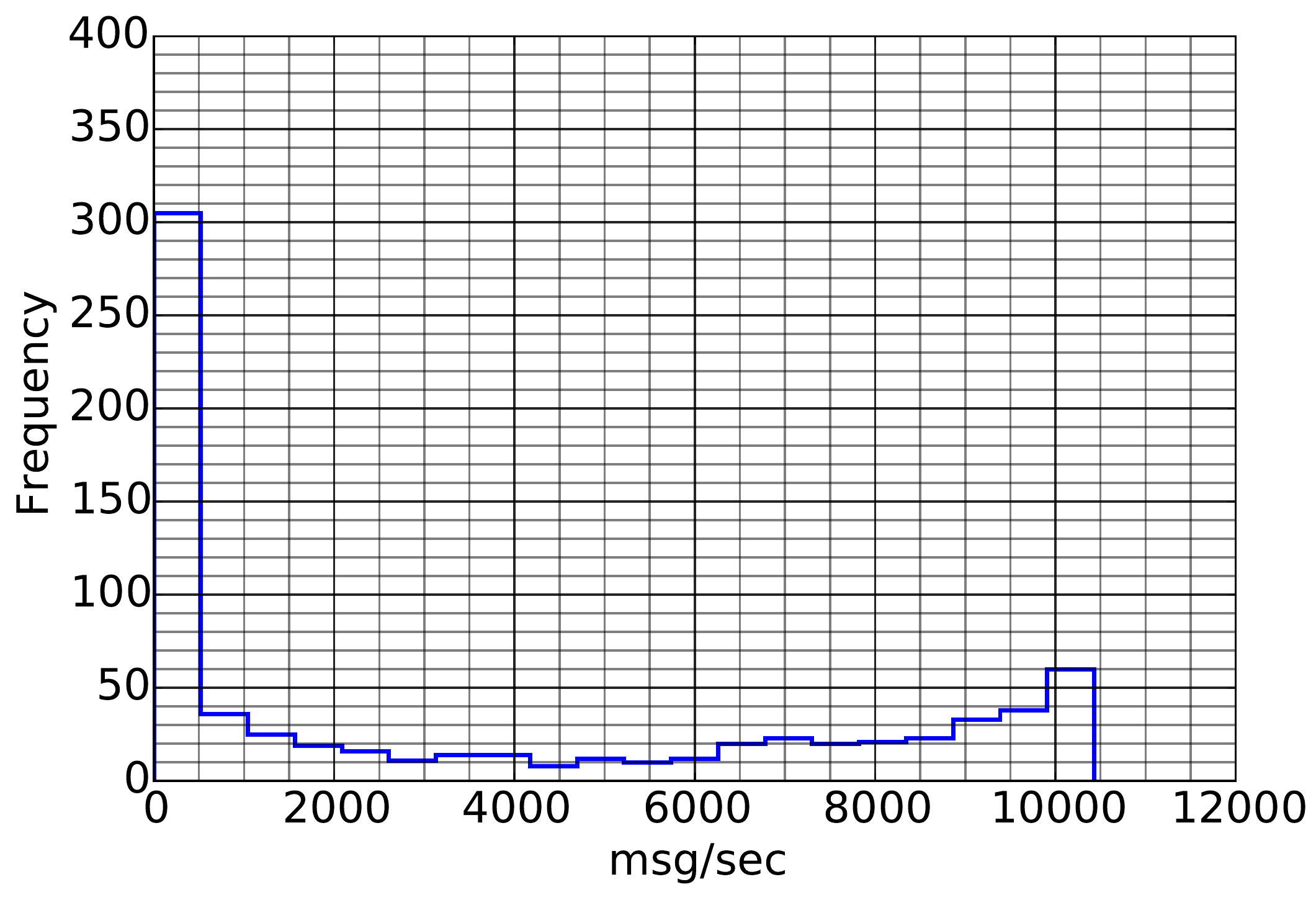}%
		\label{fig:data:energy}%
	}  
	\subfloat[FIT ]{%
		\includegraphics[width=0.3\textwidth]{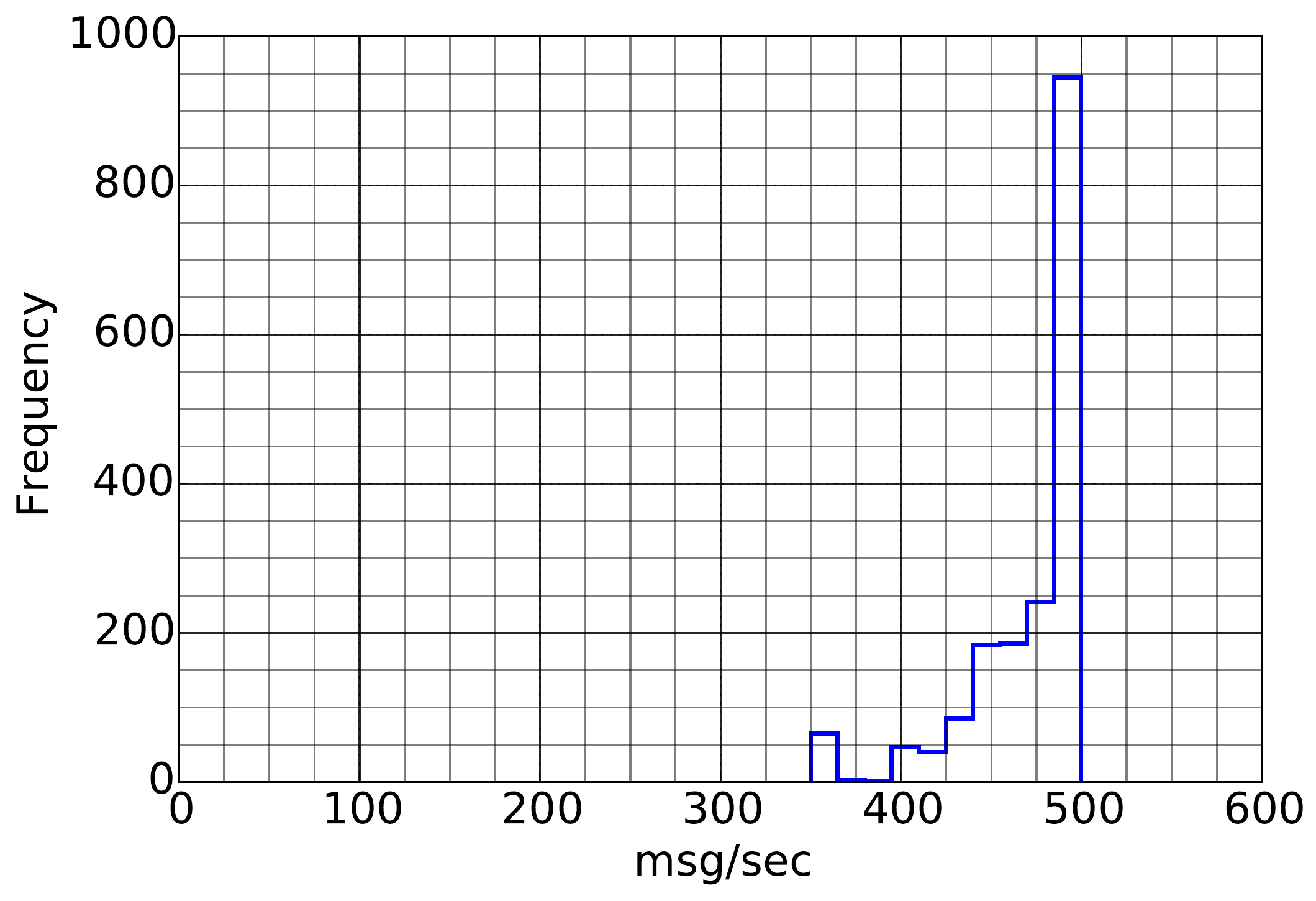}%
		\label{fig:data:health}%
	}  
	\caption{Frequency distribution of input throughput for the four workloads, with the temporal and spatial scaling used for the benchmark runs.}
	\label{fig:data-distribution}
\end{figure}

We have identified four real-world IoT data streams available in the public domain as candidates for our benchmarking workload. These correspond to domains within smart cities, which is a major contributor to the growth of IoT, taxi cab services, and personal fitness. Their features are shown in Table~\ref{tbl:datasets} and their message rate distribution is in Fig.~\ref{fig:data-distribution}.

\begin{table}[t]
	\centering
	\footnotesize
	\caption{Characteristics of Smart Cities data stream workloads used in benchmarks, with temporal and spatial scaling.}
	\resizebox{\linewidth}{!}{%
	\begin{tabular}{c|*{4}c|*{3}c|cc}
		\hline
		& \multicolumn{4}{c|}{\textbf{Raw Workload}}&\multicolumn{3}{c|}{\textbf{Scaling Factor}}&\multicolumn{2}{c}{\textbf{Effective Workload}}\\
		\textbf{Name}&\emph{Sensors}&\emph{Attrib.$^*$}  & \emph{Size} $(bytes)$ &\textbf{Distribution} &\emph{Temporal}&\emph{Spatial}&\emph{Effective} & \emph{Peak Rate} $(msg/sec)$ & \emph{Sensors}  \\ \hline\hline
		\textbf{CITY}~\cite{data:city} & 90  & 9  & 380 & Uniform & $30\times$ & $30\times$ & $900\times$ & 5,000 & 2,700 \\ \hline
		\textbf{FIT}~\cite{data:fit} & 10 & 26  & 1,024 & Uniform & $1\times$ & $1\times$ & $1\times$ & 500 & 10 \\ \hline
		\textbf{GRID}~\cite{data:grid} & 6,435 & 3  & 130  & Normal & $1\times$ & $500\times$ & $500\times$ & 10,000 & 32,17,500 \\ \hline
		\textbf{TAXI}~\cite{data:taxi} & 20,355  & 17 & 191 & Bimodal & $1,000\times$ & $1\times$ & $1,000\times$ & 4,000  & 20,355 \\ \hline \hline
		\multicolumn{10}{c}{\emph{$^*$ Every dataset has a minimum of three attributes: sensorId, timestamp and one (or more) observational field(s).}}\\
	\end{tabular}}
	\label{tbl:datasets}
\end{table}

\textbf{Sense your City (CITY)~\cite{data:city}.} 
This is an \emph{urban environmental monitoring} project~\footnote{\texttt{http://map.datacanvas.org}} that has used crowd-sourcing to deploy sensors at $7$ cities across $3$ continents in 2015, with about $12$ sensors per city. Five timestamped observations, outdoor temperature, humidity, ambient light, dust and air quality, are reported every minute by each sensor along with metadata on sensor ID and geolocation. Besides urban sensing, this also captures the vagaries of using crowd-sourcing for large IoT deployments. Data from over $2$ months is available. 
%
%
We use a single logical stream that combines the global data from all unique sensors provided in the dataset. 
Fig.~\ref{fig:data:sys} shows a narrow distribution of the message rate, with the peak frequency centered at $5,000$~msg/sec.



\textbf{NYC Taxi cab (TAXI)~\cite{data:taxi}.} 
This offers a stream of \emph{smart transportation} messages that arrive from $2M$ trips taken in 2013 on $20,355$ New York city taxis equipped with GPS. A message is generated when a taxi completes a single trip, and provides the taxi and license details, the start and end coordinates and timestamp, the distance traveled, and the cost, including the taxes and tolls paid. Other similar transportation datasets are also available~\footnote{\texttt{https://github.com/fivethirtyeight/uber-tlc-foil-response}}, though we chose ours based on the richness of the fields. 
This data has a bi-modal event rate distribution that reflects the morning and evening commutes, with peaks at $300$ and $3,200$~events/sec. We use 7~days of data from 14-Jan-2013 to 20-Jan-2013 for our benchmark runs. 

\textbf{Energy dataset (GRID)~\cite{data:grid}.}
This is a univariate dataset that reports the energy consumption for each smart meter in a pilot smart grid deployment in Ireland. The actual dataset had $6,435$ unique sensors and emits a reading every half an hour. 
Data is available from over 500~days of observations. It shows a normal distribution of data around each half an hour timestamp. The final peak rate of dataset used in the experimental runs is $10,000$~events/sec.

\textbf{MHealth dataset (FIT)~\cite{data:fit}.} 
The MHEALTH (Mobile HEALTH) dataset consists of body motion and vital signs recordings for ten volunteers of diverse profiles collected when performing physical activities. Sensors measure in different parts of the subject's body 
collect acceleration, rate of turn, magnetic field, and ECG data, among others, at a constant rate of $50~Hz$.  
We the merge 10 subjects data into a single global stream, with messages having the subject ID as sensor ID. It has a constant rate of $500$~events/sec as shown in Fig.~\ref{fig:data:health}. 

While these datasets correspond to real values collected from the domain, they are representative samples from even larger datasets that are typically proprietary. In order to capture the real scale of these data streams, we make use of temporal and spatial scaling. 
\emph{Temporal scaling} allows us to accelerate the data rate by time-compressing messages that were generated over a longer interval into a smaller one. 
For e.g., when the CITY data is temporally scaled by $30\times$, its original rate per sensor goes from an average of $6$~msg/min to $180$~msg/min, and $7$ days of wall-clock time get reduces to $336$~mins of benchmark time. This causes the shape of the distribution in Fig.~\ref{fig:data-distribution} to be retained, but widens the X Axis. Temporal scaling is relevant when the raw workload data that is available is not representative of the sampling rates that are expected in contemporary IoT sensors and domains. Considering that GPS sensors placed in taxis report their location each second for navigation and monitoring rather than only at the end of the trip, we use a temporal scaling factor of $1000\times$ for the TAXI workload. 

\emph{Spatial scaling}, on the other hand, allow us to simulate a larger number of sensors than available in the raw data. 
This is necessary when data streams are available only from a small sample of sensors. Here, we consider data streams from the same sensor but during different time windows (e.g., days) to act as if they are from different sensors but at a precious time. This too does not affect the shape of the message rate distribution, but expands the Y Axis. For e.g., tn the GRID data, a spatial scaling of $500\times$ increases the $6,435$ smart meters present in original dataset to $3,217,500$ unique meters, which is more representative of a city-scale deployment. 
Similarly, a $30\times$ spatial scaling in CITY (in addition to the temporal scaling) causes the 336~mins of benchmark time to further reduce to 12~mins of benchmark time, while 
increasing the sensor count from $90$ to $2,700$.  

These two scaling factors are also shown in Table~\ref{tbl:datasets}, along with the effective number of sensors and the peak rate after applying these factors. As we can see, using scaling to create workloads that are representative of real-world scenarios also achieves diversity in the event rate distribution profiles for the input streams, and also offers peak rates that span from 500 to 10,000~msg/sec.

\section{Evaluation of Proposed Benchmarks}
\label{sec:results}

\subsection{Benchmark Implementation}
We implement the 27 micro-benchmarks as generic Java tasks that can consume and produce objects. These tasks are building blocks that can be composed into micro-dataflows and the ETL, STATS, PRED and TRAIN dataflows using any DSPS that is being benchmarked. To validate our proposed benchmark, we compose these dataflows on the Apache Storm open source DSPS, popular for fast-data processing, using its Java APIs. We then run these for the four stream workloads and evaluate them based on the metrics we have defined. The benchmark is available online at \texttt{\url{https://github.com/dream-lab/riot-bench}}.

In Storm, each task logic is wrapped by a \emph{bolt} that invokes the task for each incoming tuple and emits zero or more response tuples downstream. 
The dataflow is composed as a \emph{topology} that defines the edges between the bolts, and the \emph{groupings} which determine duplicate or hash semantics. 
We have implemented a scalable data-parallel event generator that acts as a source task (\emph{spout}). It loads time-series tuples from a given SenML file and replays them as an input stream to the dataflow. While the spatial scaling of the workloads is performed offline as a pre-processing step, our generator can perform temporal scaling online, as it emits the message.  
We generate random integers as tuples at maximum rate for the micro-benchmarks, and replay the original datasets by scaling their native rates as in table~\ref{tbl:datasets} for the application benchmarks, following the earlier distribution. 

\subsection{Experimental Setup}
We use Apache Storm $1.0.1$ running on OpenJDK 1.7, and hosted on \emph{Ubuntu 14.0} Virtual Machines (VMs) in the Southeast Asia data center of Microsoft Azure public cloud. For the micro-benchmarks, Storm runs the task being benchmarked on one exclusive \texttt{D1} size VM ($1$ Intel Xeon E5-2660 core at 2.2~GHz, $3.5$~GiB RAM, $50$~GiB SSD), while the supporting source and sink tasks and the master service run on a \texttt{D4} size VM ($8$ Intel Xeon E5-2660 core at 2.2~GHz cores, $28$~GiB RAM, $400$~GiB SSD). 
The larger VM for the supporting tasks and services ensures that they are not the bottleneck, and helps benchmark the peak rate supported by the micro-benchmark task on a single core VM.

For the ETL, STATS, TRAIN and PRED application benchmarks, we use \texttt{D3}~VMs ($4$ Intel Xeon E5-2660 core at 2.2~GHz cores, $14$~GiB RAM, $200$~GiB SSD) for all the tasks of the dataflow, while reserving additional \texttt{D4}~VMs to exclusively run the source and sink tasks, and the Storm master service. Storm requires the users to explicitly assign the data parallelism per task, and the total number of resources in the cluster. We determine the number of cores and data parallelism required by each task using a simple resource allocation algorithm, as follows. 

We have the peak rate supported by the single-threaded task on a single core as given by the micro-benchmarks, and the peak rate seen for that task for a given application and stream workload by examining the dataflow and selectivity. For tasks where the expected rate in the dataflow is less than its peak rate supported on one core, we assign it an exclusive core and two threads. In cases that are I/O bound (e.g., MQTT, Azure storage) rather than CPU bound, we require multiple task instances on a single core to leverage data parallelism, and sometimes multiple cores as well. Table ~\ref{table:slots} shows the number of cores and VMs assigned for running the experiments with the applications and stream workloads.

\begin{table}[t]
	\centering
	\footnotesize
	\caption{The number of resources assigned, given as ``cores, VMs'', for each application benchmark and workload. Each VM has 4 cores.}
	\label{table:slots}
	\begin{tabular}{c|c|c|c|c}
		\hline
		\textbf{App.}&\qquad \textbf{CITY}\qquad\qquad &\qquad \textbf{FIT}\qquad\qquad &\qquad \textbf{GRID}\qquad\qquad &\qquad \textbf{TAXI}\qquad\qquad \\ \hline\hline
		\textbf{ETL}   & 11, 3 & 8, 2  & 14, 3 & 10, 3 \\ \hline
		\textbf{STATS} & 27, 7 & 10, 3 & 11, 3 & 32, 8     \\ \hline
		\textbf{TRAIN} & 7, 2  & 7, 2  & N/A$^\dagger$   & 7, 2  \\ \hline
		\textbf{PRED}  & 10, 3 & 9, 3  & N/A$^\dagger$   & 9, 3  \\ \hline
		\hline
		\multicolumn{5}{c}{\emph{$^\dagger$ Benchmarks are not done for the particular applications with the GRID dataset as}}\\
		\multicolumn{5}{c}{\emph{it is univariate and DTC and MLR tasks require multiple fields.}}
	\end{tabular}
\end{table}

We log the ID and timestamp for each message at the source and the sink tasks in-memory to calculate the latency, throughput and jitter metrics. 
We also sample the CPU and memory usage on all VMs every 5~secs to plot the utilization metrics. Each experiment runs for $\sim 10$~mins of wallclock time.


\subsection{Micro-benchmark Results}
\begin{figure}[t]
	\centering
	\subfloat[Peak Throughput]{
		\includegraphics[width=0.40\textwidth]{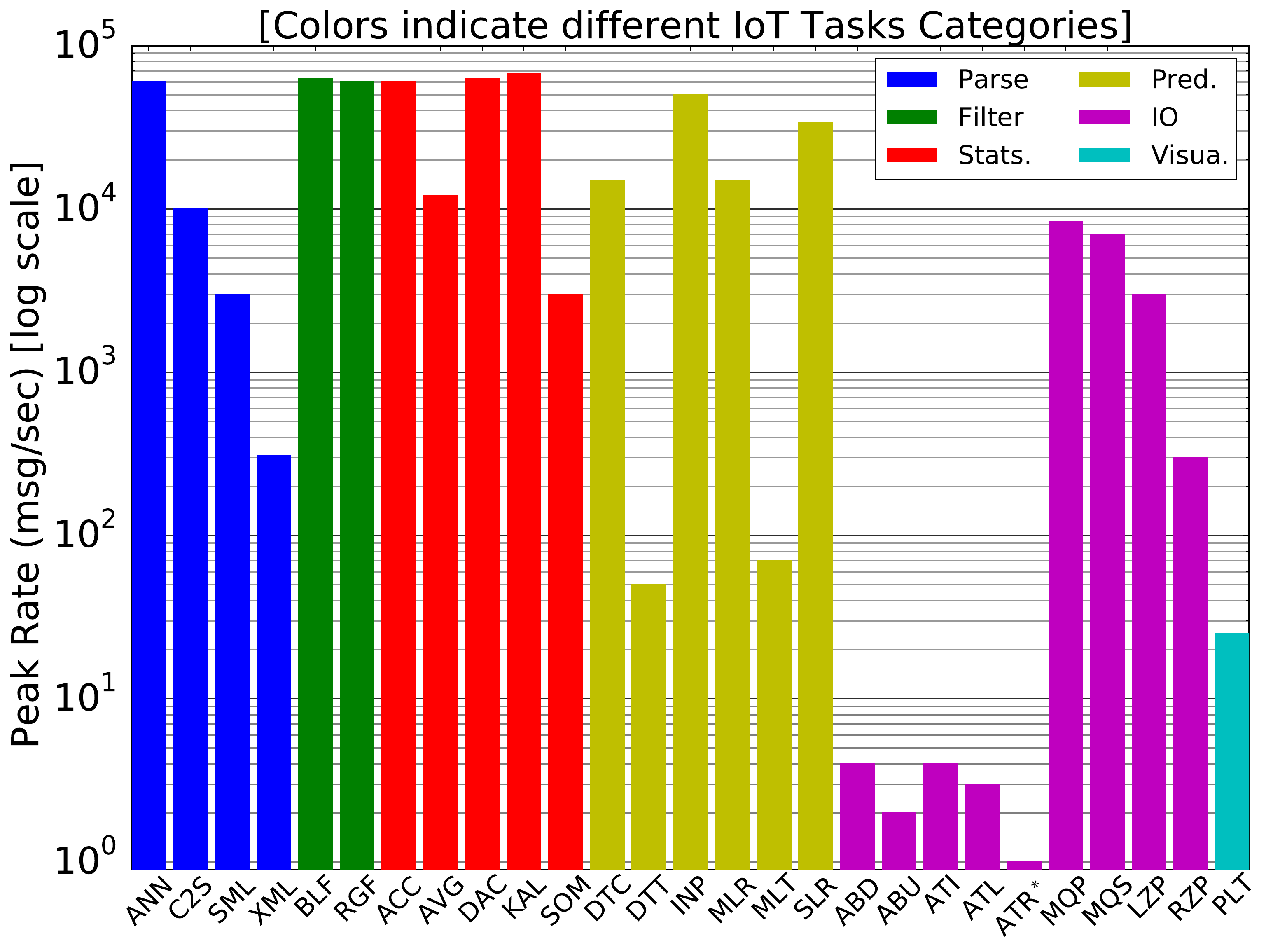}
		\label{fig:storm:micro:peakthru}
	}\\
	\subfloat[End-to-end latency]{
		\includegraphics[width=0.40\textwidth]{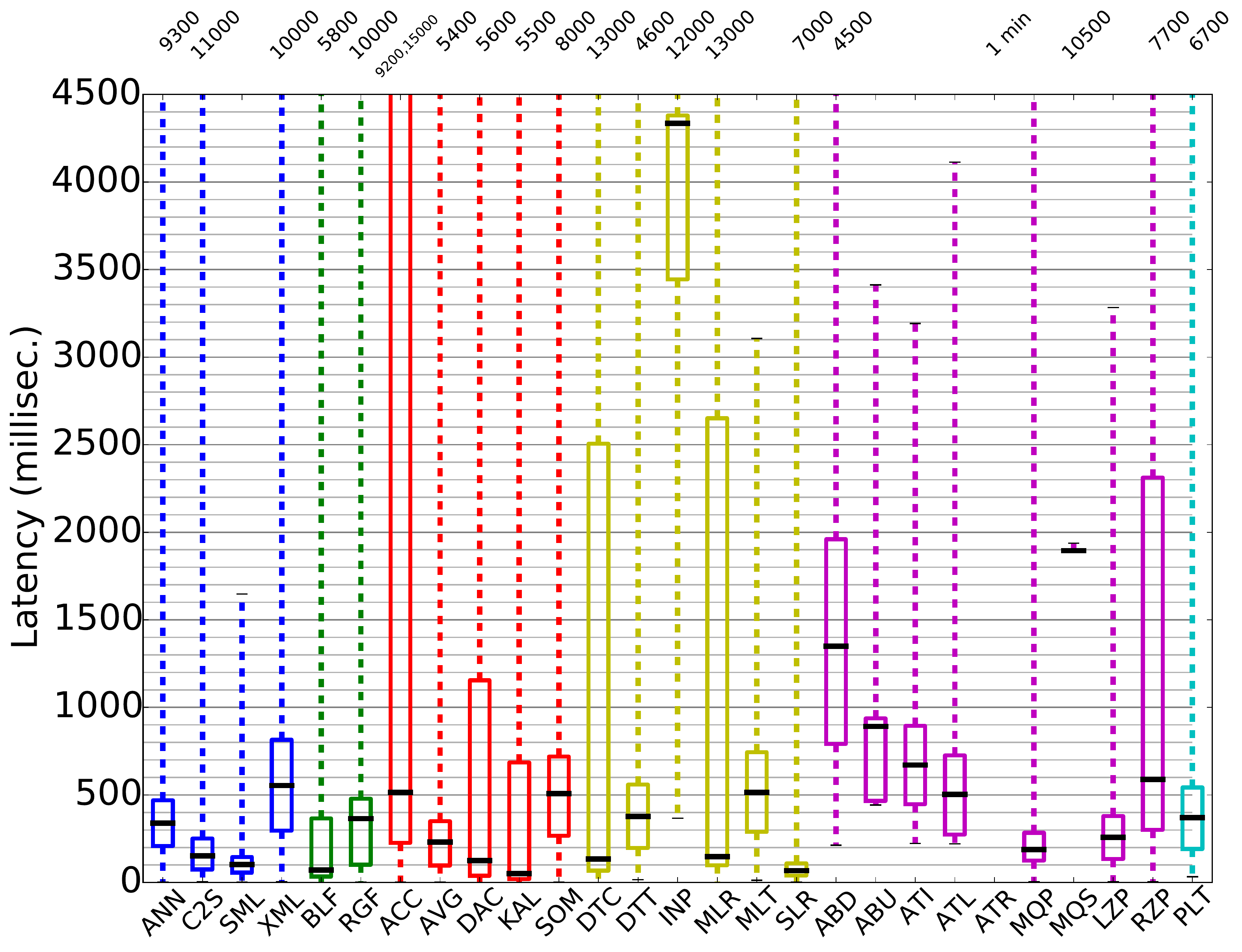}
		\label{fig:storm:micro:latency}
	}
	\subfloat[Jitter]{
		\includegraphics[width=0.40\textwidth]{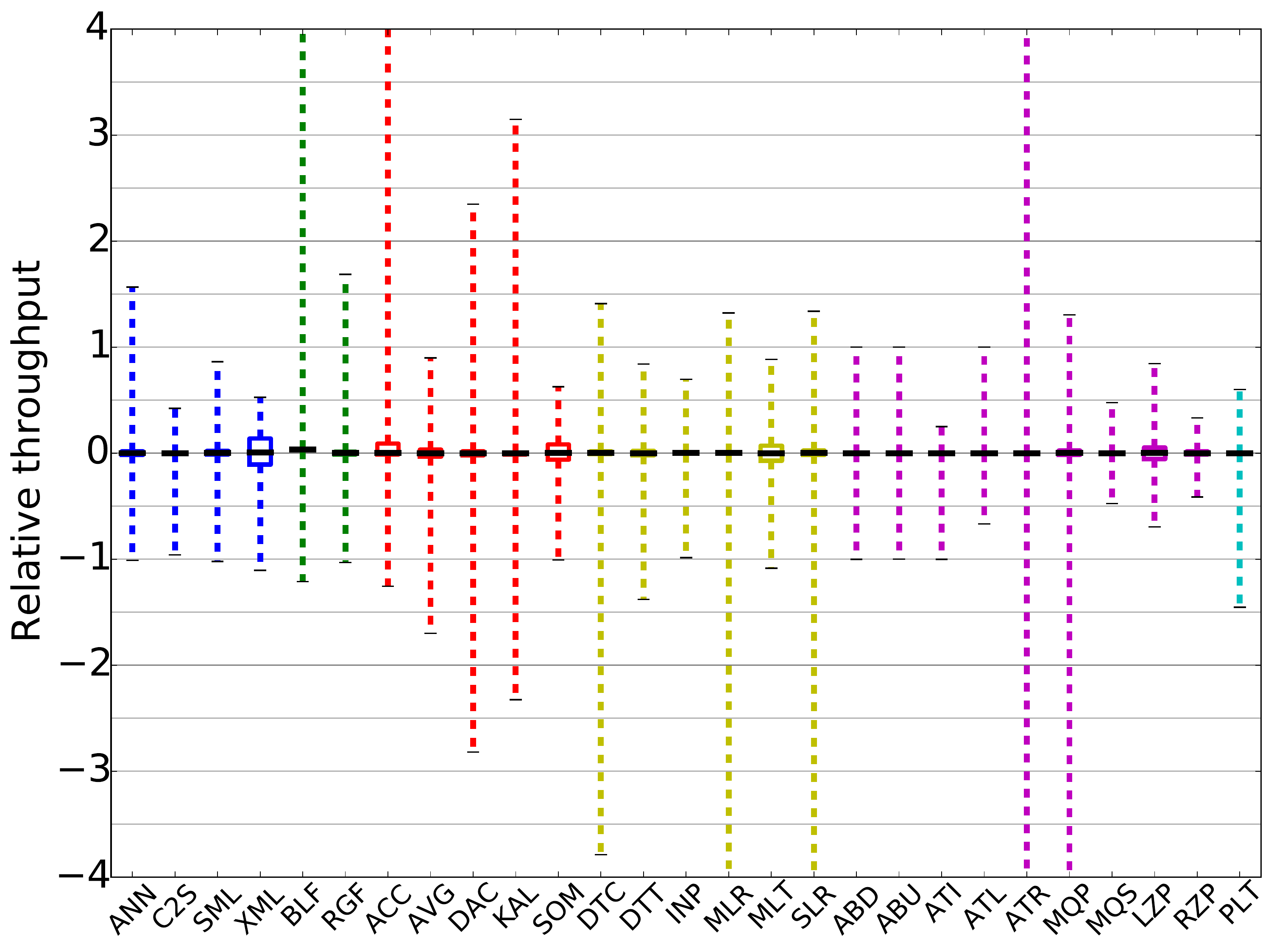}
		\label{fig:storm:micro:jitter}
	}\\
	\subfloat[CPU\%]{
		\includegraphics[width=0.40\textwidth]{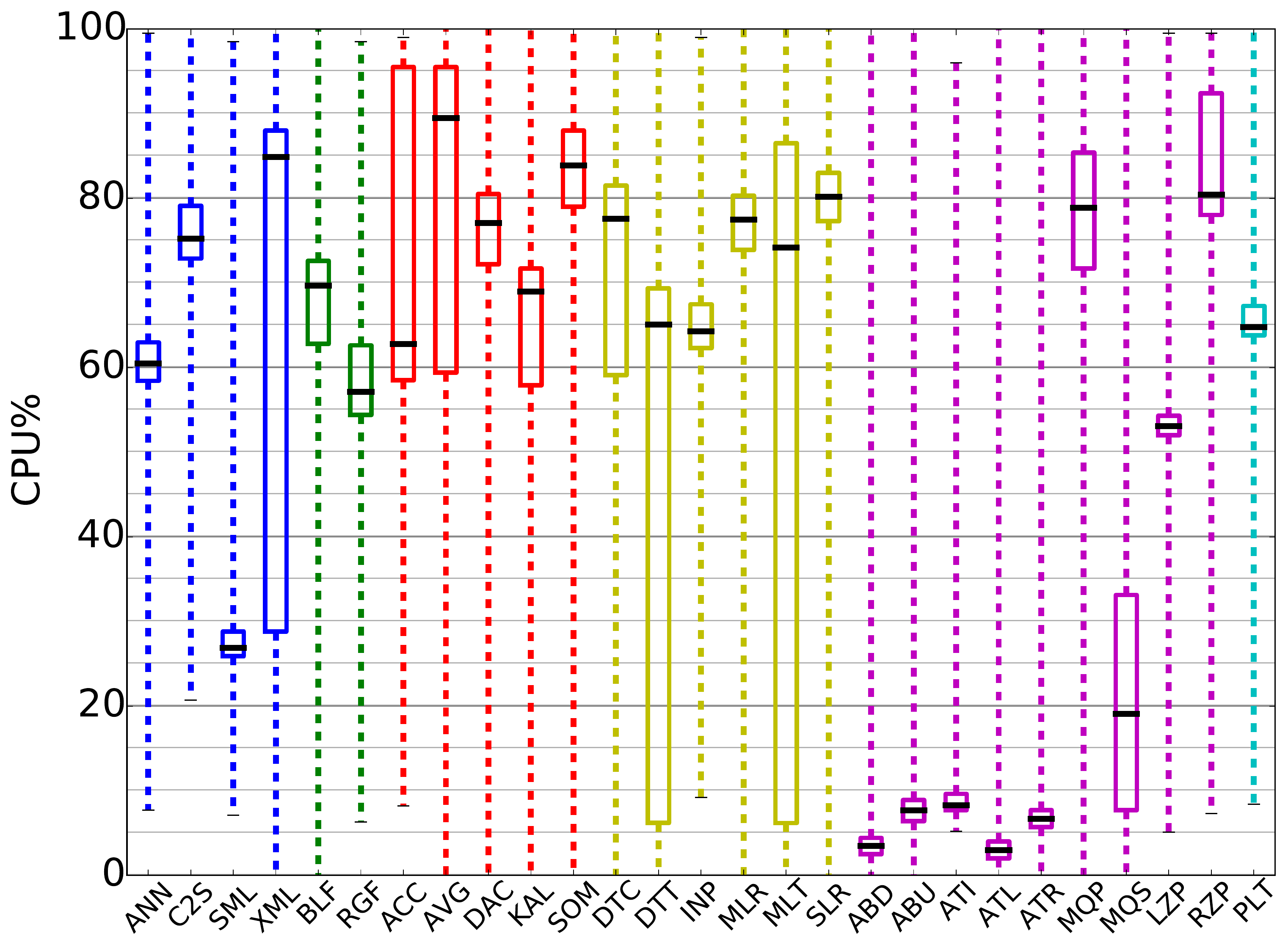}
		\label{fig:storm:micro:cpu}
	} 
	\subfloat[MEM\%]{
		\includegraphics[width=0.40\textwidth]{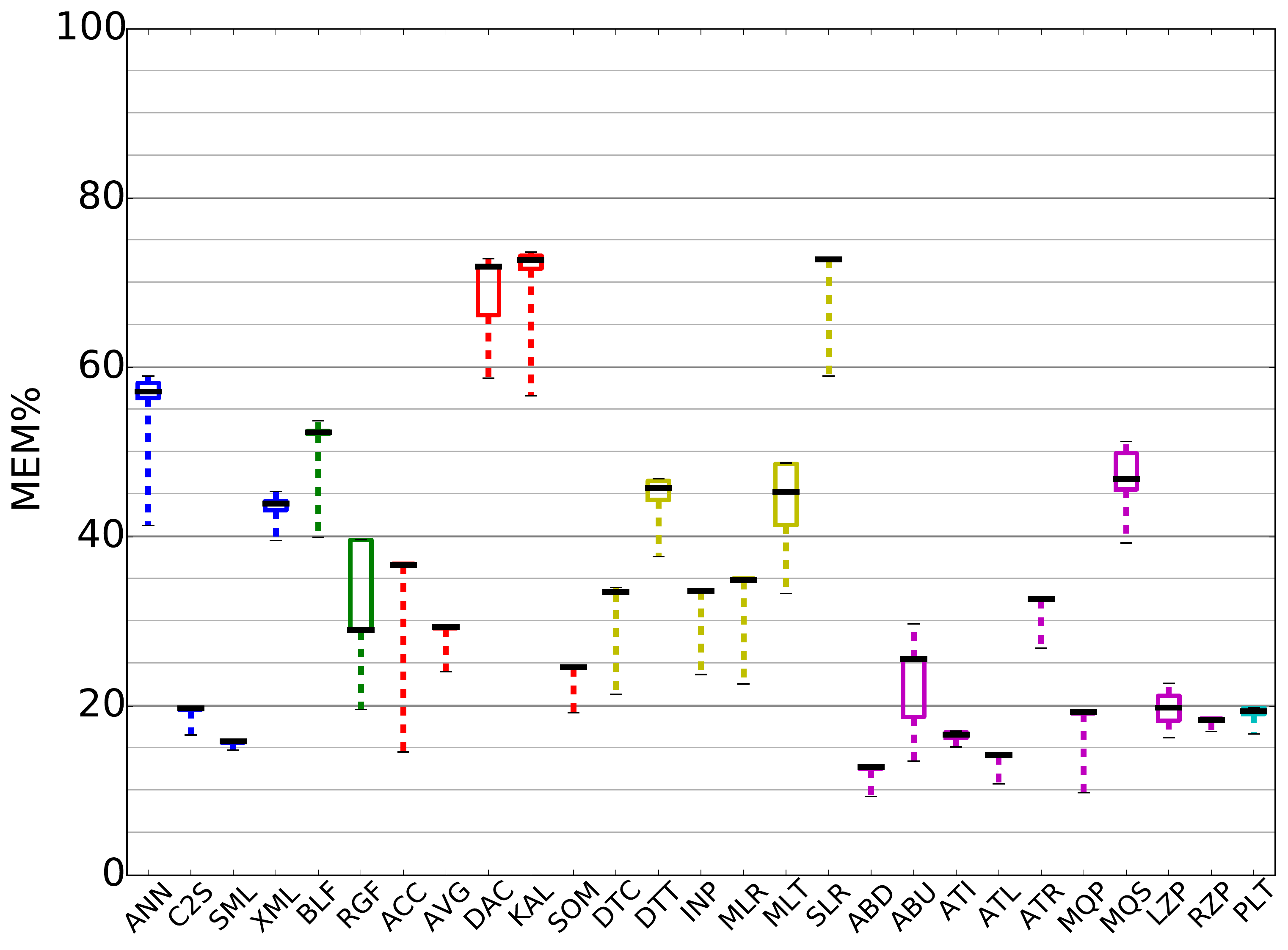}
		\label{fig:storm:micro:mem}
	}
	\caption{Performance of micro-benchmark tasks for integer input stream at peak rate.}
	\label{fig:storm:micro:bm}
\end{figure}

Fig.~\ref{fig:storm:micro:bm} shows plots of the different metrics evaluated for the micro-benchmark tasks on Storm when running at their peak input rate supported on a single \texttt{D1} VM with one thread. The \emph{peak sustained throughput} per task is shown in Fig.~\ref{fig:storm:micro:peakthru} in \emph{log-scale}. We see that most tasks can support $3,000$~msg/sec or higher rate, going up to $68,000$~msg/sec for ANN, BLF, RGF, ACC, DAC and KAL. XML parsing is highly CPU bound and has a peak throughput of only $310~msg/sec$. SML parse supports much higher rate than XML with less CPU usage, indicating that it is a better fit for streaming IoT applications than the XML format. 
DTT and MLT uses WEKA library for model training and supports only $50$ and $70~msg/sec$ rate, CPU being the bottleneck. PLT uses the XChart~\cite{xchart:lib} Java charting library and supports only $25~msg/sec$ rate as it is CPU intensive around $70\%$ as shown in~\ref{fig:storm:micro:cpu} at peak rate. 

The Azure operations are I/O bound on the Cloud service and slow due to the web service latency. ATR supports only $1~msg/min$, as the task has to scan the full table on Azure with, e.g., $753,382$ records for the Taxi dataset, to query over non-key attributes on single Azure table partition. Better input rates can be achieved by storing Azure table on multiple partitions with query attributes as partition or row-key. RZP supports $300~msg/sec$ while LZP supports $3,000~msg/sec$ -- RZP has to write the Zip file to a remote shared directory while LZP uses a local disk. 

The inverse of the peak sustained throughput gives the \emph{mean latency}, and we do not explicitly plot it. However, it is interesting to examine the \emph{end-to-end latency}, calculated as the time taken between emitting a message from the source, having it pass through the benchmarked task, and arrive at the sink task. This is the effective time contributed to the total tuple latency by this task running within Storm, including framework overheads. We see that while the mean latencies should be in sub-milliseconds for the observed throughputs, the box plot for end-to-end latency (Fig.~\ref{fig:storm:micro:latency}) varies widely up to $2,600~ms$ for Q3, except ACC and INP task. This wide variability could be because of non-uniform task execution times due to which slow executions queue up incoming messages that suffer higher queuing time, such as for DTC and MLR that both use the WEKA library. Or tasks supporting a high input rate in the order of $10,000~msg/sec$, such as DAC and KAL, may be more sensitive to even small per-tuple overhead of the framework, say, caused by thread contention between the Storm system and worker threads, or queue synchronization. 

The Azure tasks that have a lower throughput also have a higher end-to-end latency, but much of which is attributable directly to the task latency. ATR has a latency of $1~min$ due to scanning of the large table. ACC shows wide distribution of latency due to variability in the complexity of operation performed on it. Events associated with a single sensor ID are stored in a time-ordered queue until the threshold count is reached, upon which it extracts all the accumulated values and passes it downstream. MQS shows latency of $1,900~ms$ with no whiskers as the task logic just polls a local queue of messages being populated by the subscribed messages arriving from the broker.

The box-plot for \emph{jitter} (Fig.~\ref{fig:storm:micro:jitter}) shows values close to zero in all cases. This indicates the long-term stability of Storm in processing the tasks at peak rate, without unsustainable queuing of input messages. The wider whiskers indicate the occasional mismatch between the expected and observed output rates. ATR again has a high range for the whiskers as its rate is very low at $1~msg/min$; thus even minor variation in rate shows high jitter values.

The box plots for CPU utilization (Fig.~\ref{fig:storm:micro:cpu}) shows the single-core VM effectively used at $70\%$ or above in all cases except for the SML, MQS and Azure tasks that are I/O bound. MQS is bounded by the number of threads as single thread is busy in polling the message queue which is not CPU intensive. SML is having low CPU of $\approx{30\%}$, the reason being that as we are using a JSON representation for SenML which is less CPU intensive as compared to XML.
The memory utilization (Fig.~\ref{fig:storm:micro:mem}) appears to be higher for tasks that support a high throughput, potentially indicating the memory consumed by messages waiting in queue rather than consumed by the task logic itself. MQS shows a high memory usage ($\approx{50\%}$) even for a low rate due to buffering of incoming messages from the broker in a queue that is asynchronously being polled. 
Similarly, memory for DTT and MLT is $\approx{45\%}$ because a batch of nearly thousand rows is stored in memory for model training triggered by every incoming input message. 


\subsection{Application Results}

\begin{figure}[t]
	\centering
	\subfloat[ETL]{
		\includegraphics[width=0.17\textwidth]{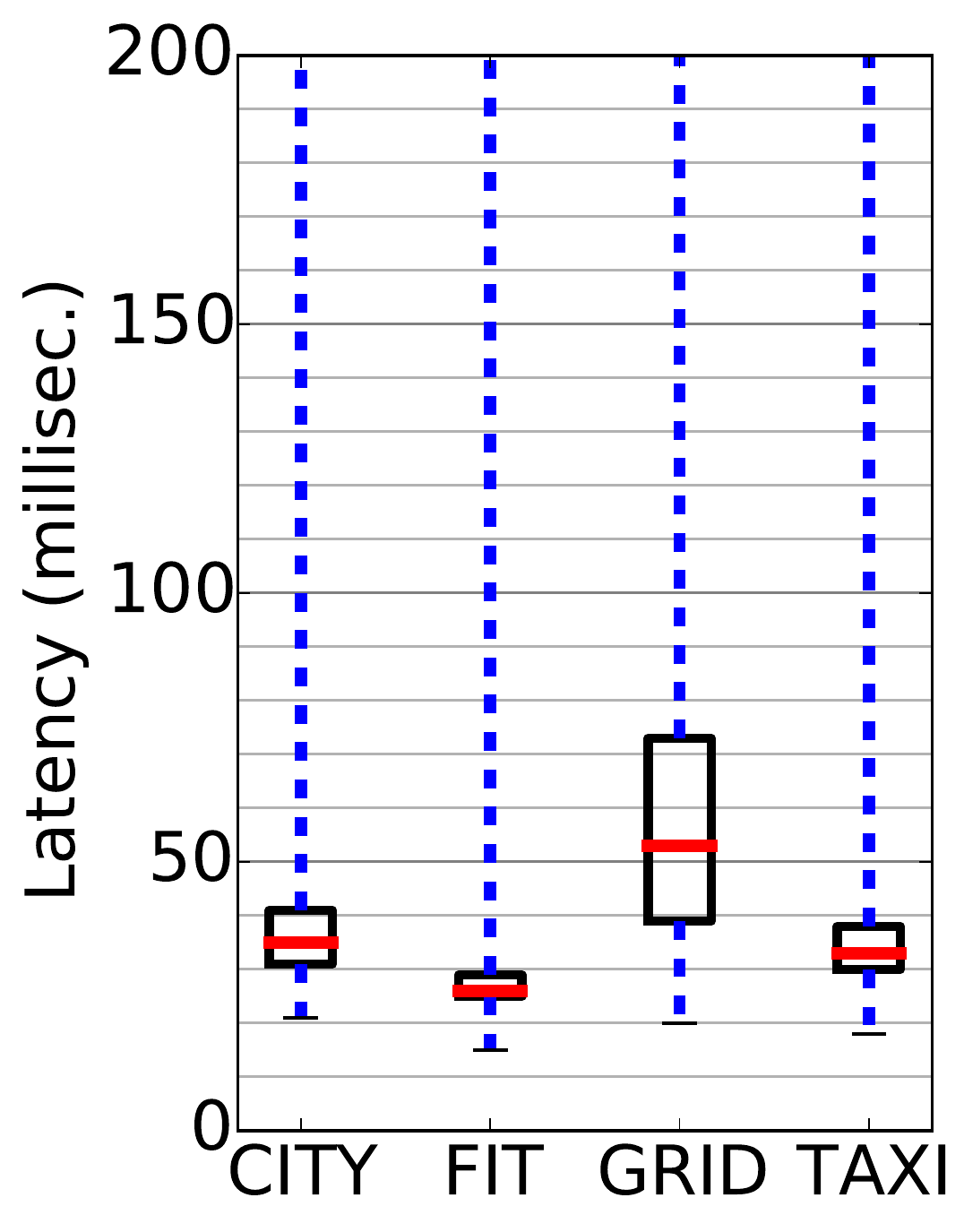}
		\label{fig:storm:etl:latency}
	}
	\subfloat[STATS]{
		\includegraphics[width=0.17\textwidth]{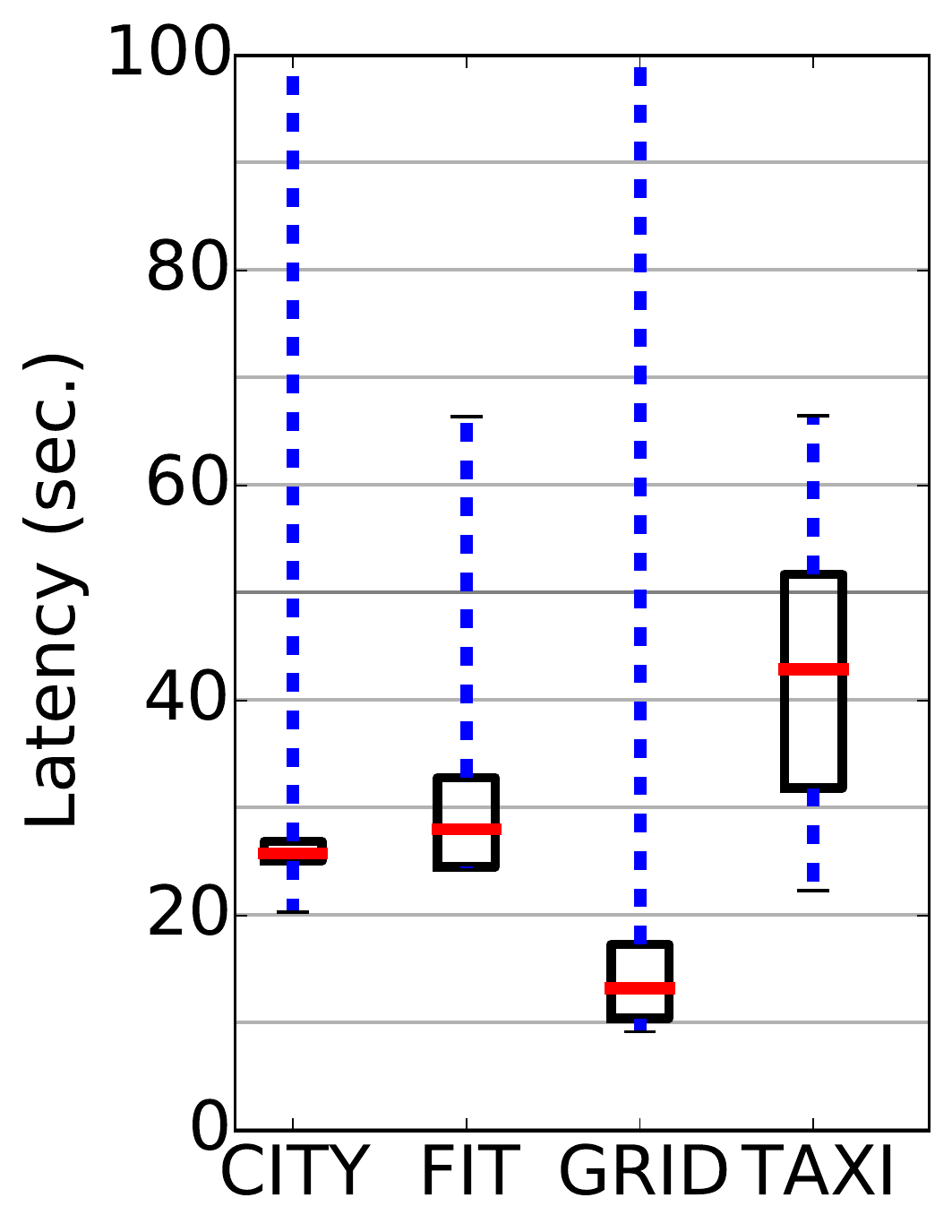}
		\label{fig:storm:stats:latency}
	}
	\subfloat[TRAIN]{
		\includegraphics[width=0.17\textwidth]{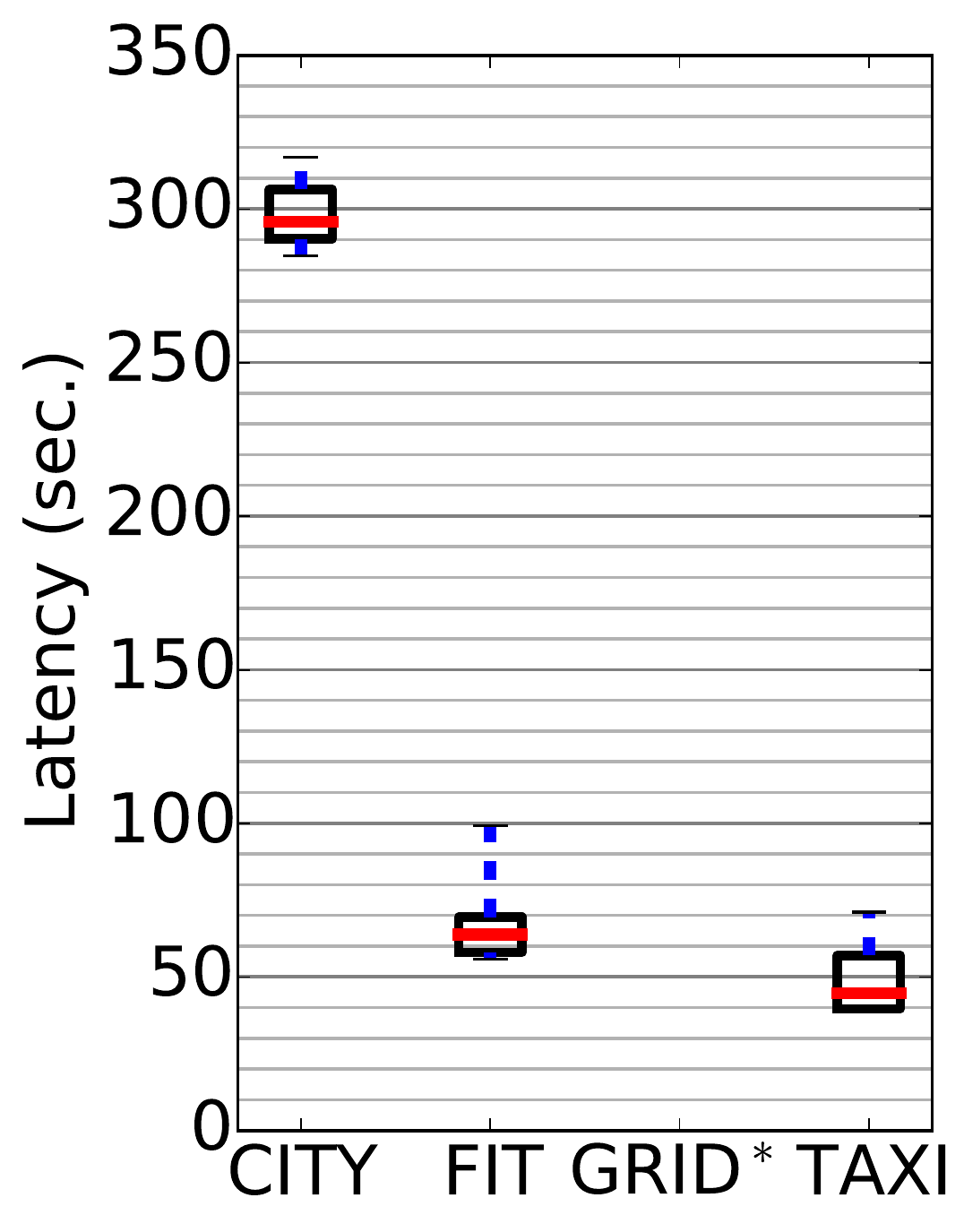}
		\label{fig:storm:train:latency}
	}
	\subfloat[PRED]{
		\includegraphics[width=0.17\textwidth]{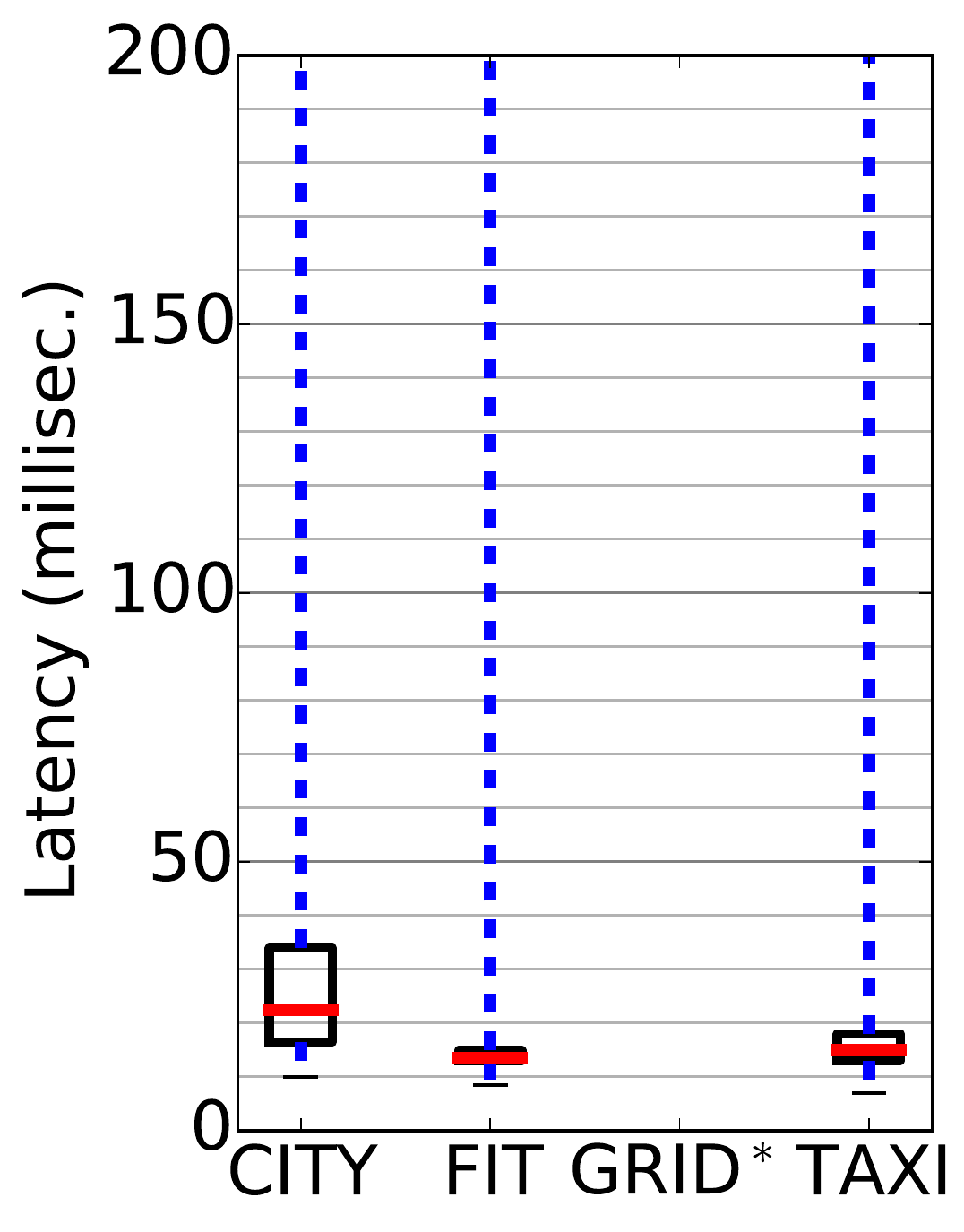}
		\label{fig:storm:pred:latency}
	}
	\caption{End-to-end latency plots for application benchmarks on workloads. ETL and PRED are in millisec and STATS and TRAIN are in sec. *TRAIN and PRED are not run for GRID workload as it has only the target field, and no additional field to predict upon.}
	
\end{figure}
\begin{figure}[t]
	
	\centering
	\subfloat[ETL]{
		\includegraphics[width=0.25\textwidth]{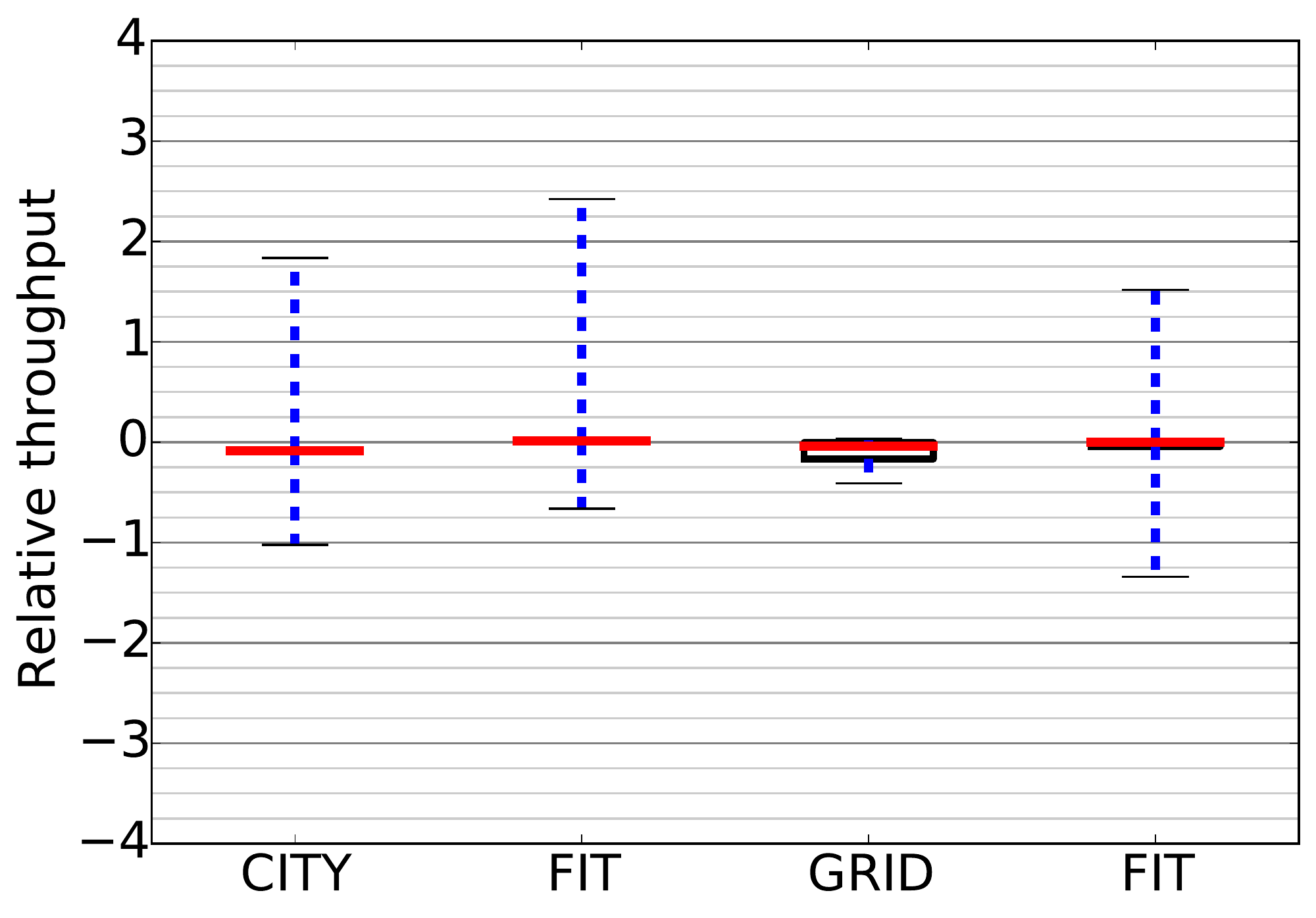}
		\label{fig:storm:app:etl:jitter}
	}
	\subfloat[STATS]{
		\includegraphics[width=0.25\textwidth]{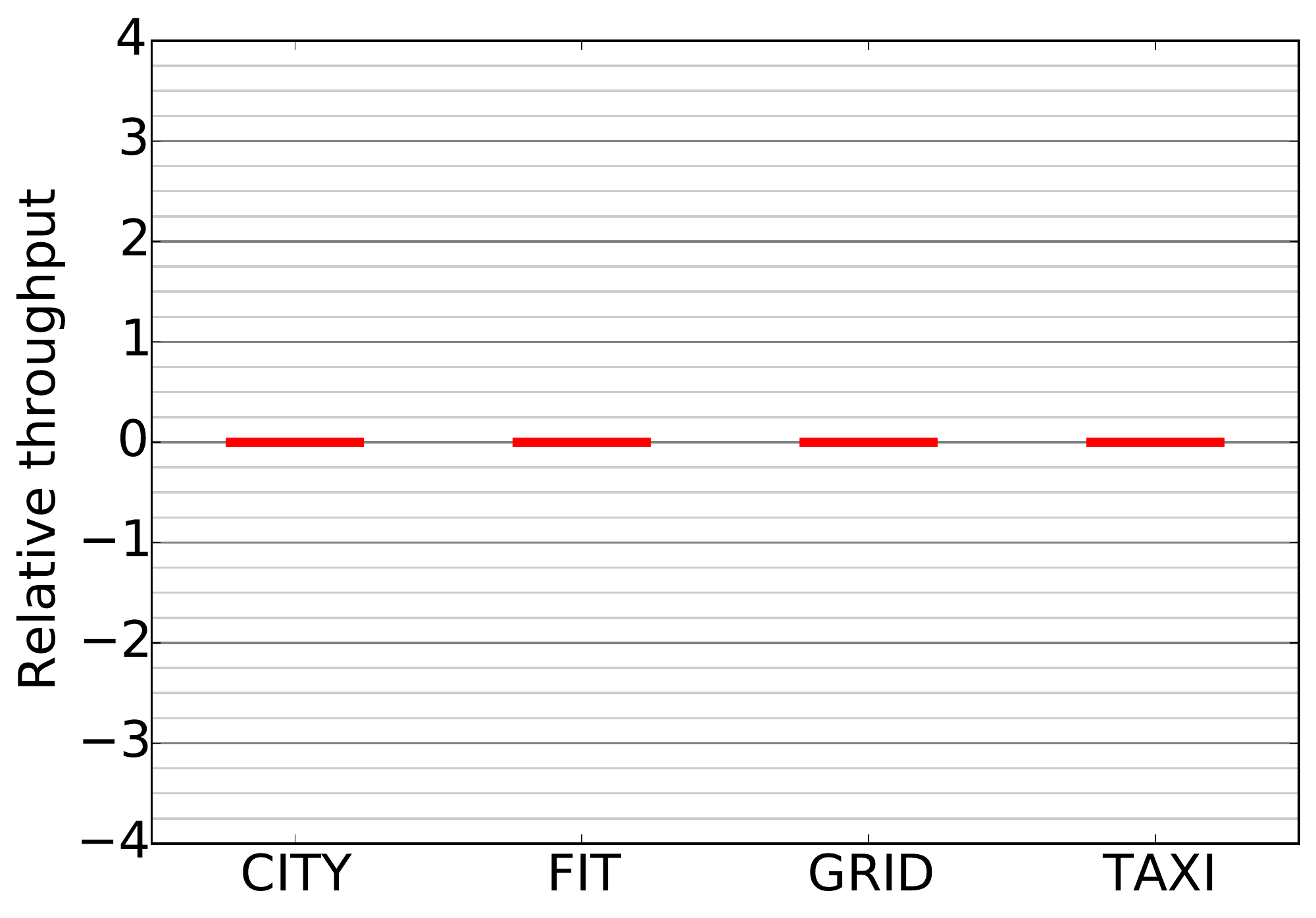}
		\label{fig:storm:app:stats:jitter}
	} 
	\subfloat[TRAIN]{
		\includegraphics[width=0.25\textwidth]{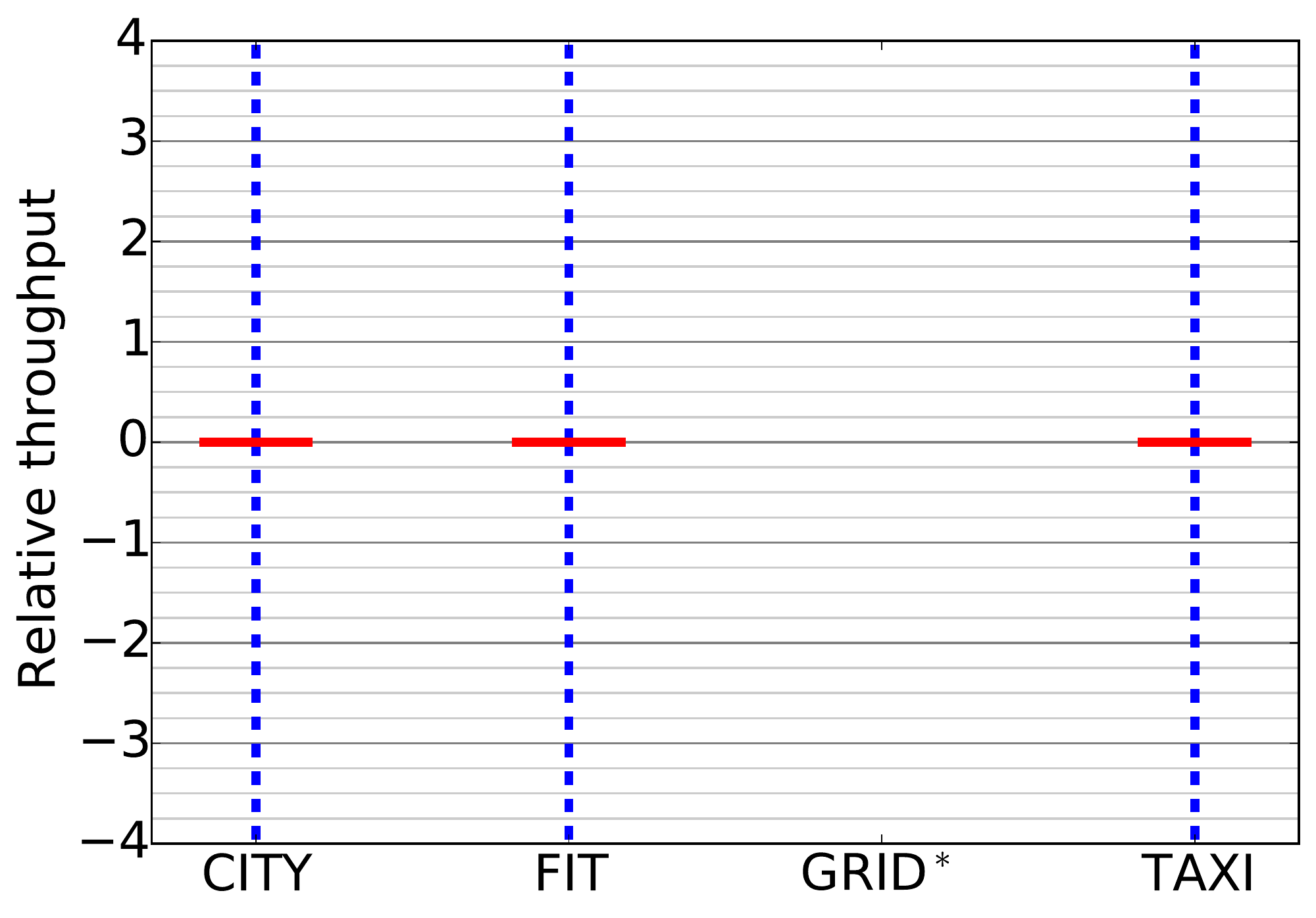}
		\label{fig:storm:app:train:jitter}
	} 
	\subfloat[PRED]{
		\includegraphics[width=0.25\textwidth]{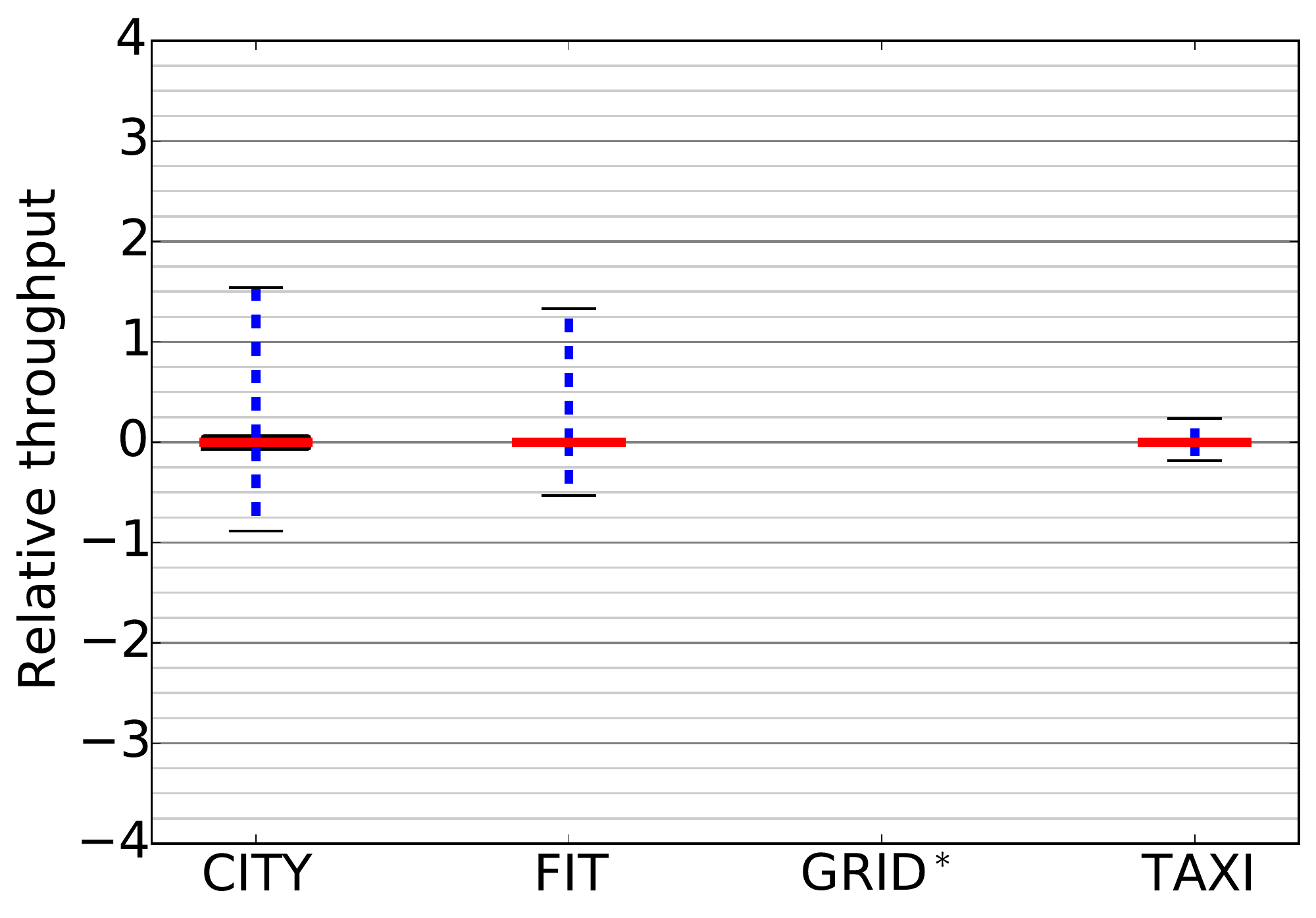}
		\label{fig:storm:app:pred:jitter}
	}
	\caption{Jitter plots for application benchmarks on workloads. *TRAIN and PRED are not run for GRID workload as it has only the target field, and no additional field to predict upon.} 
	\label{fig:storm:app:jitter}
\end{figure}
\begin{figure}[t]
	
	\subfloat[CITY]{
		\includegraphics[width=0.235\textwidth]{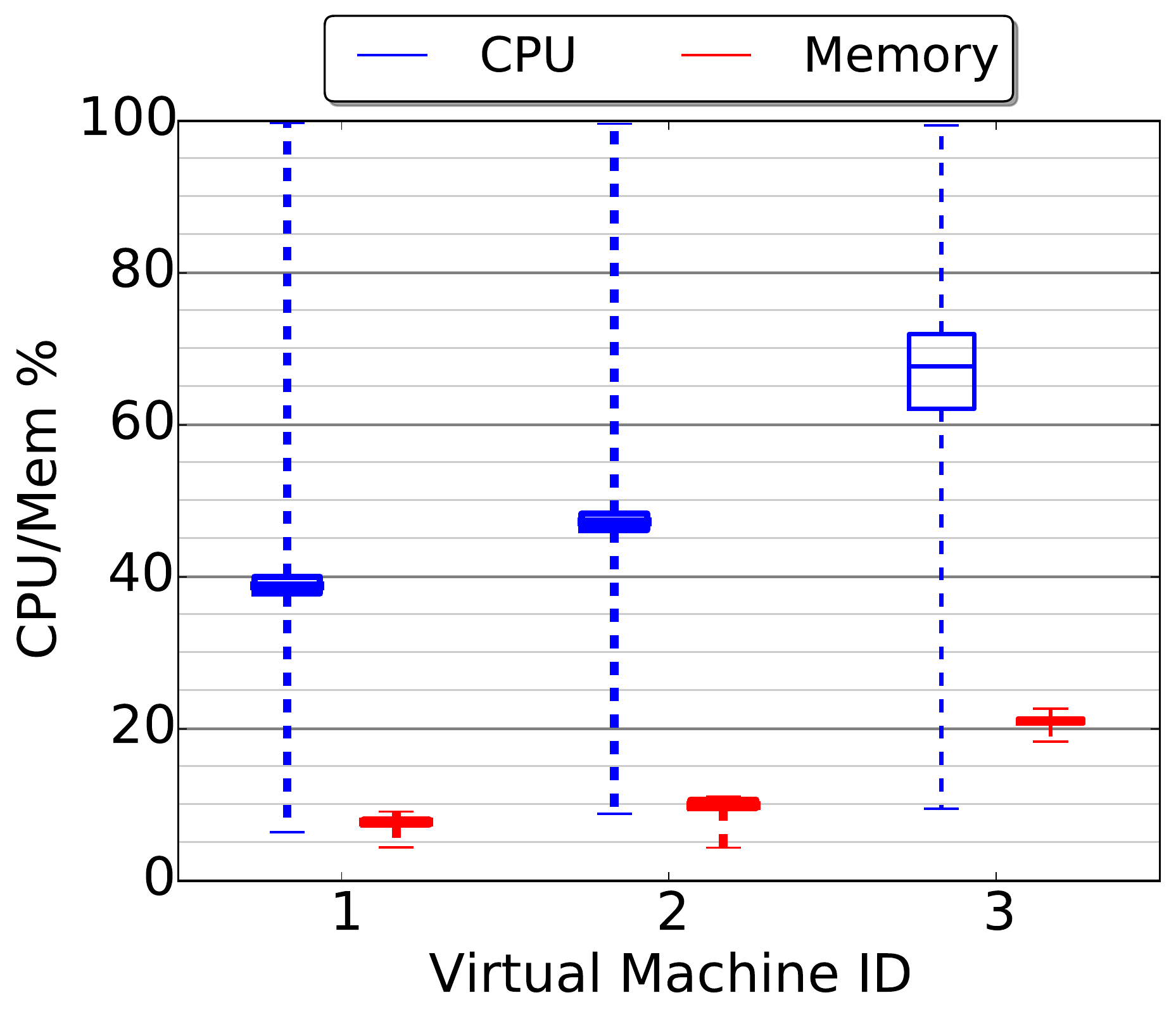}
		\label{fig:storm:etl:city:cpu}
	}
	\subfloat[FIT]{
		\includegraphics[width=0.235\textwidth]{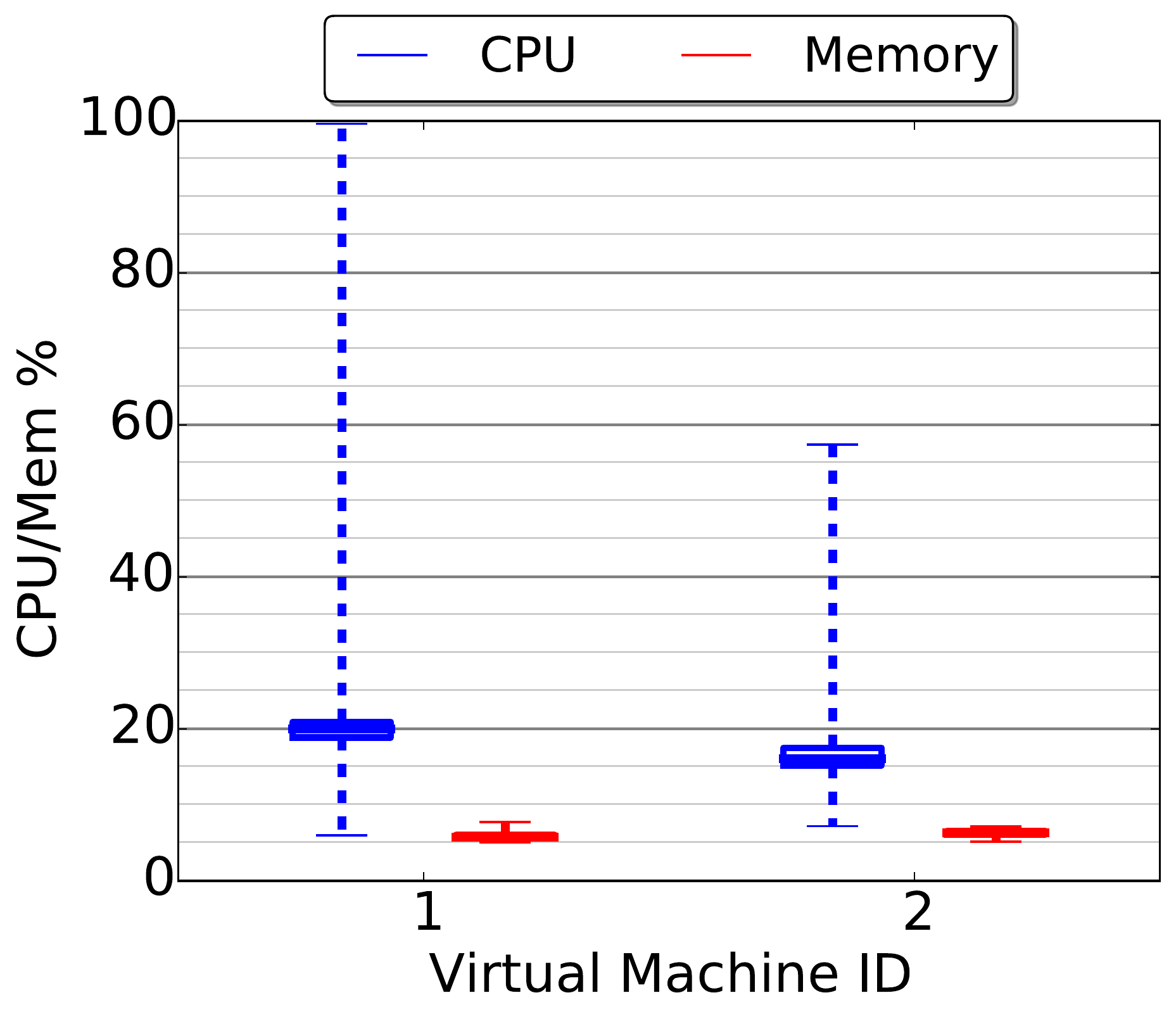}
		\label{fig:storm:etl:fit:cpu}
	}
	\subfloat[GRID]{
		\includegraphics[width=0.235\textwidth]{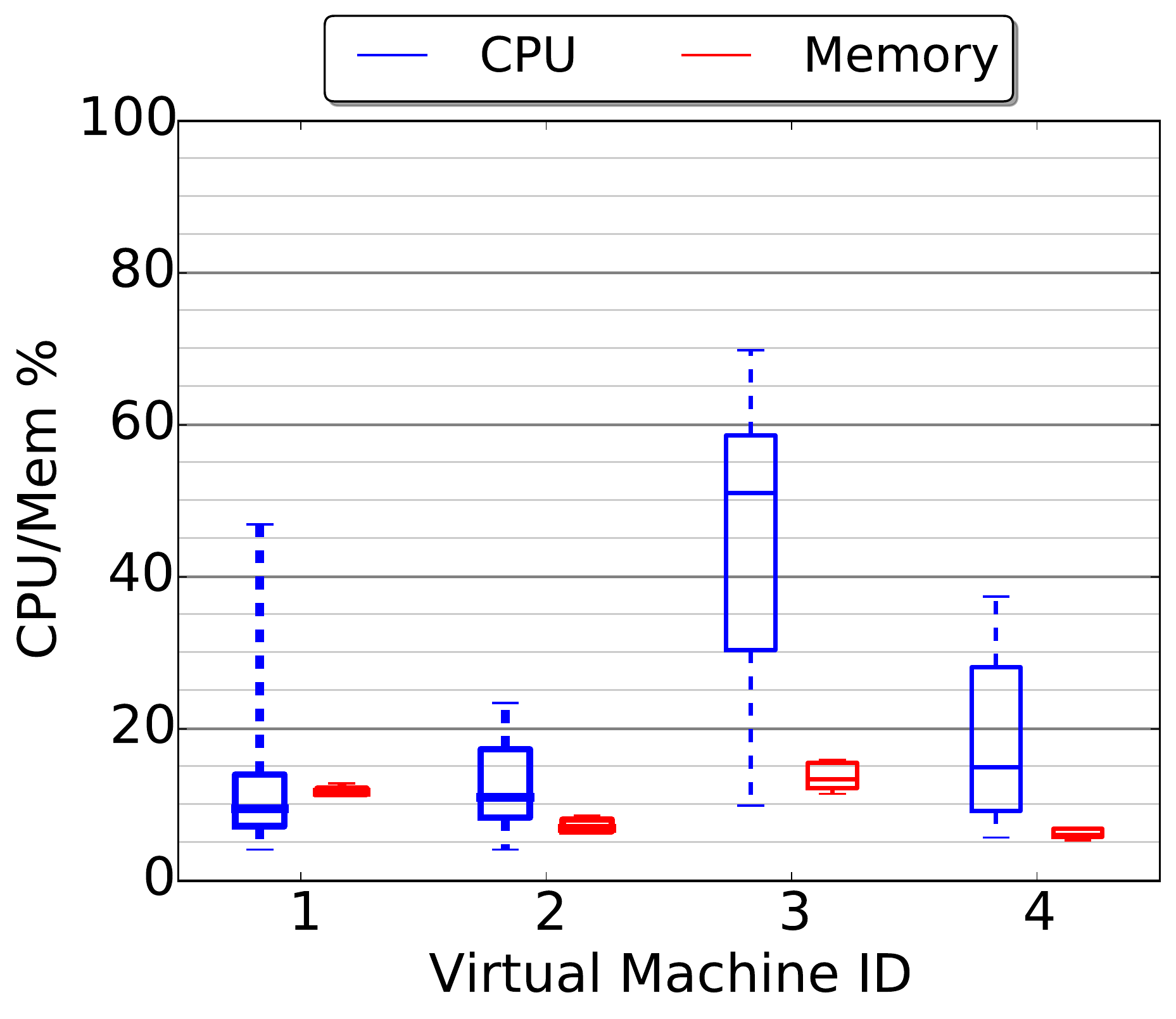}
		\label{fig:storm:etl:grid:cpu}
	}
	\subfloat[TAXI]{
		\includegraphics[width=0.235\textwidth]{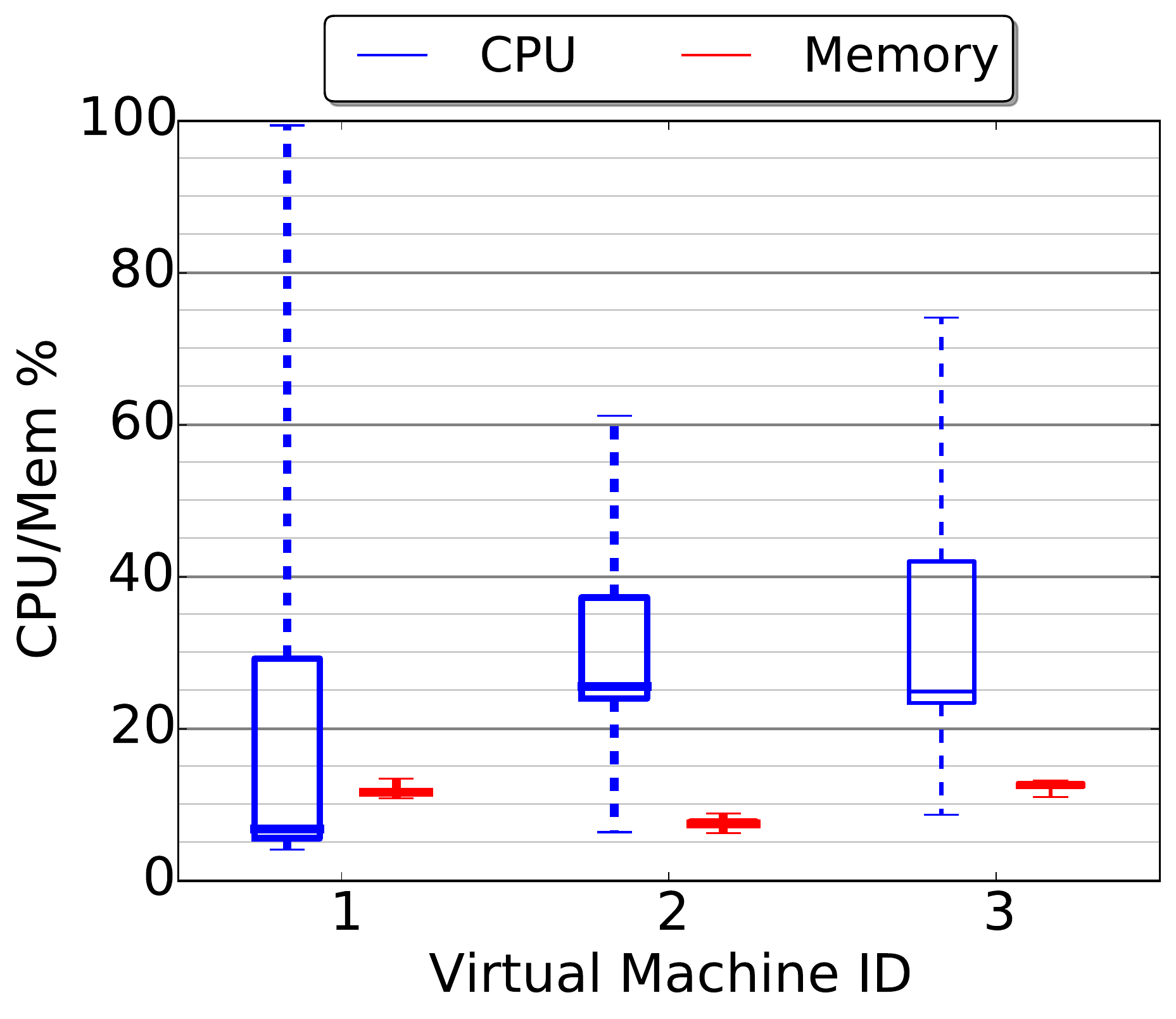}
		\label{fig:storm:etl:taxi:cpu}
	}
	\caption{CPU and Memory utilization plots for \emph{ETL} application benchmark on all workloads.}
	\label{fig:storm:etl:cpu}
\end{figure}

\begin{figure}[t]
	\subfloat[CITY]{
		\includegraphics[width=0.235\textwidth]{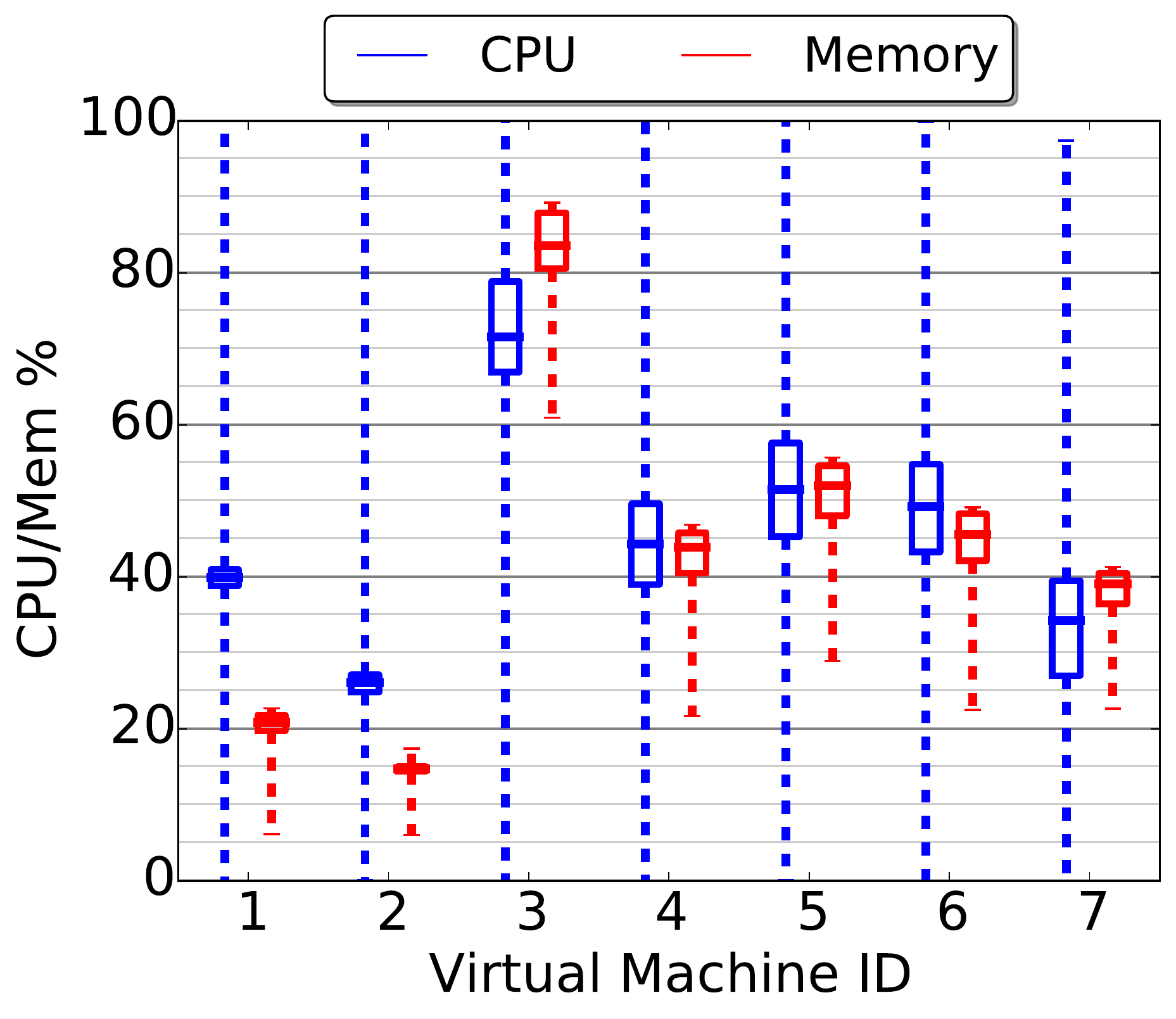}
		\label{fig:storm:stats:city:cpu}
	}
	\subfloat[FIT]{
		\includegraphics[width=0.235\textwidth]{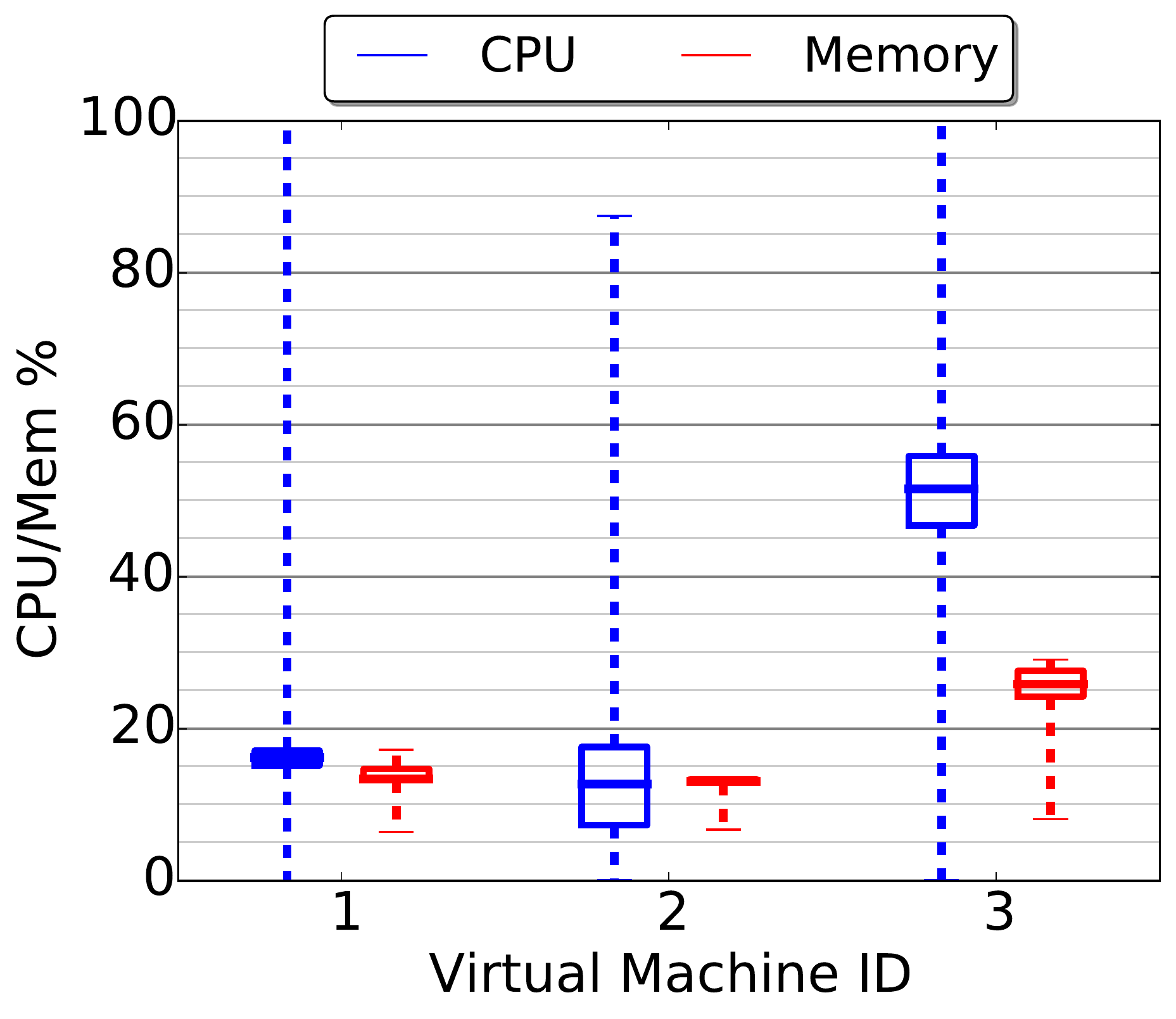}
		\label{fig:storm:stats:fit:cpu}
	}
	\subfloat[GRID]{
		\includegraphics[width=0.235\textwidth]{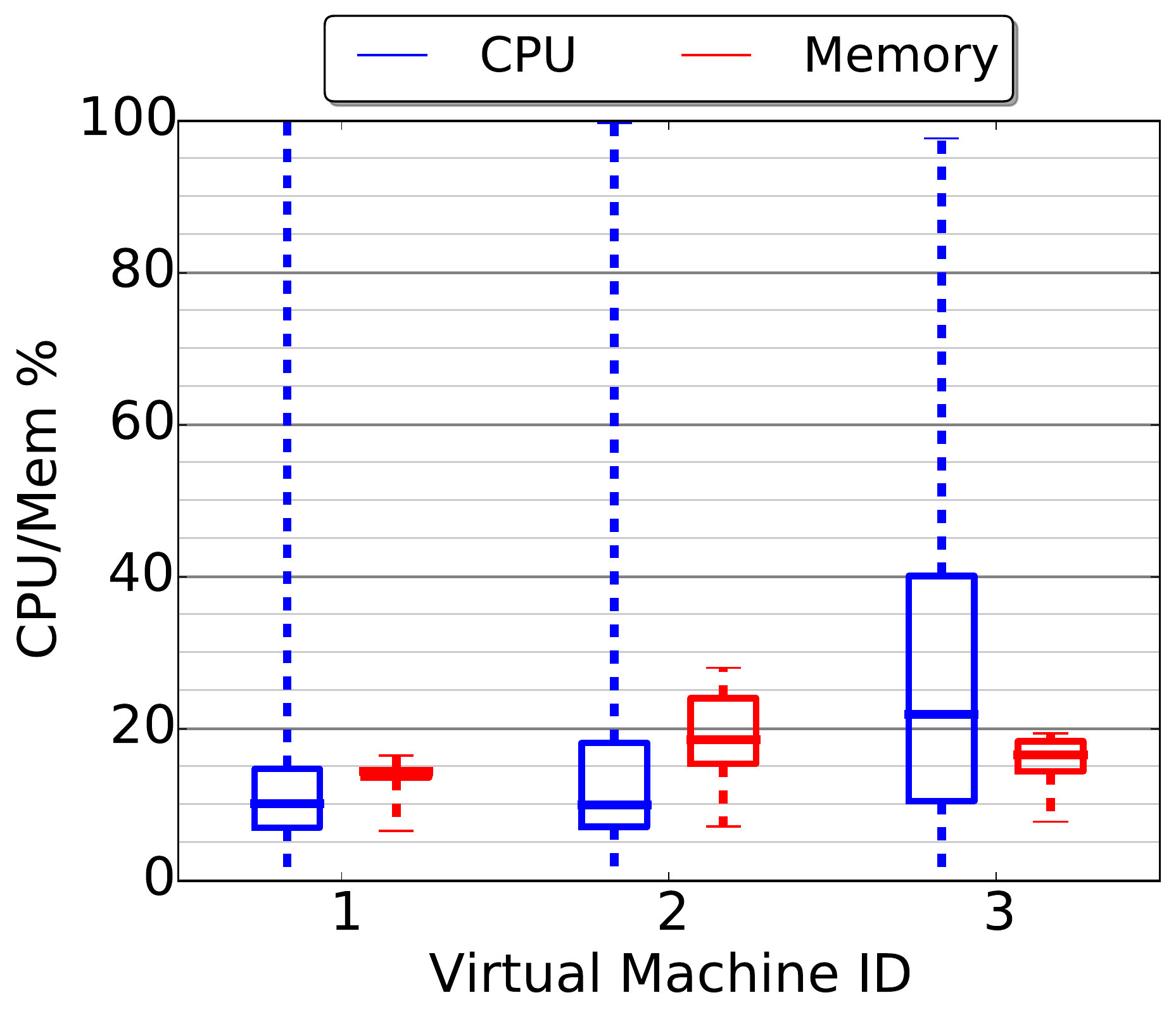}
		\label{fig:storm:stats:grid:cpu}
	}
	\subfloat[TAXI]{
		\includegraphics[width=0.235\textwidth]{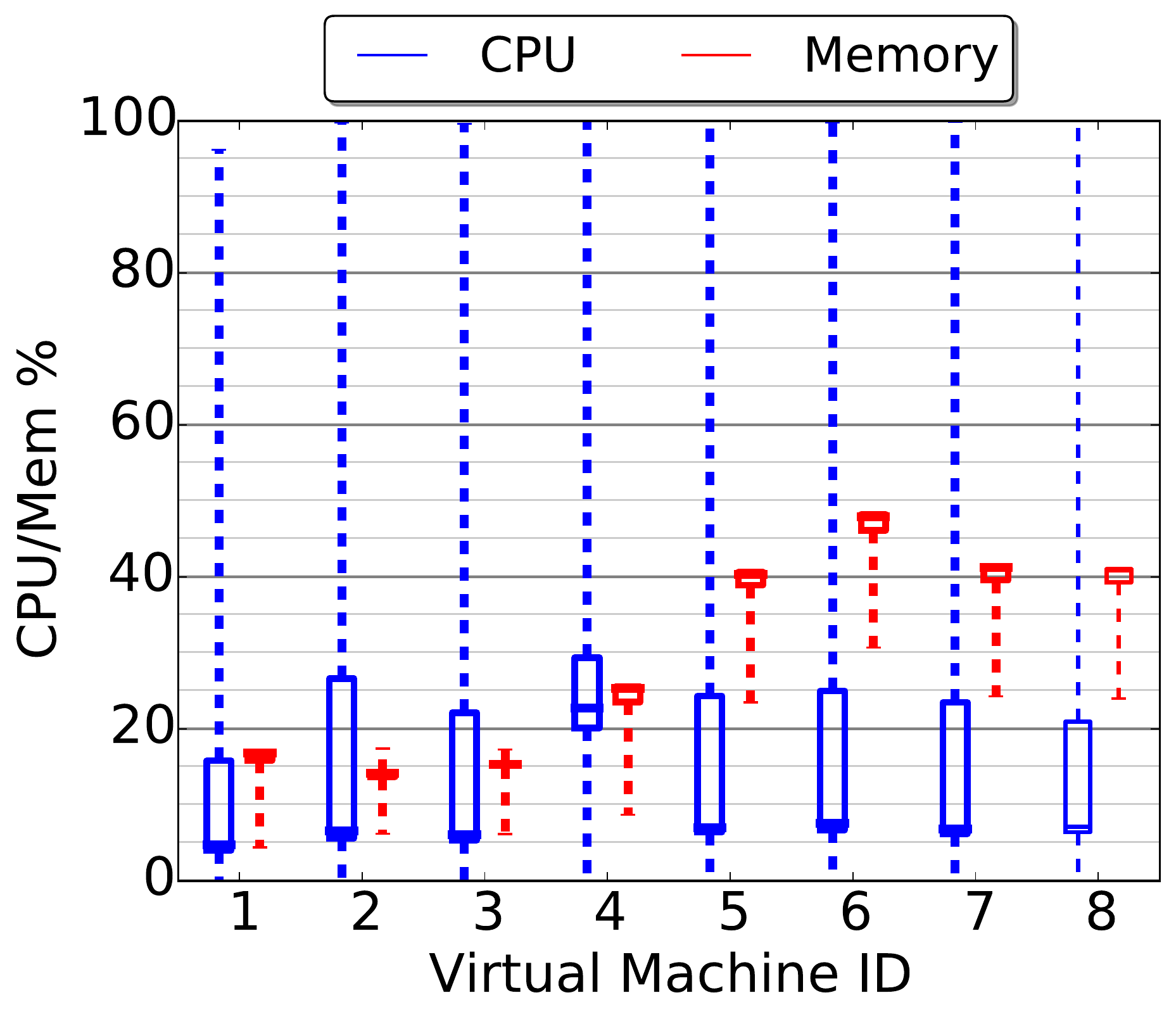}
		\label{fig:storm:stats:taxi:cpu}
	}
	\caption{CPU and Memory utilization plots for \emph{STATS} application benchmarks on all workloads.}
	\label{fig:storm:stats:cpu}
\end{figure}

\begin{figure}[t]
	\centering
	\subfloat[CITY]{
		\includegraphics[width=0.235\textwidth]{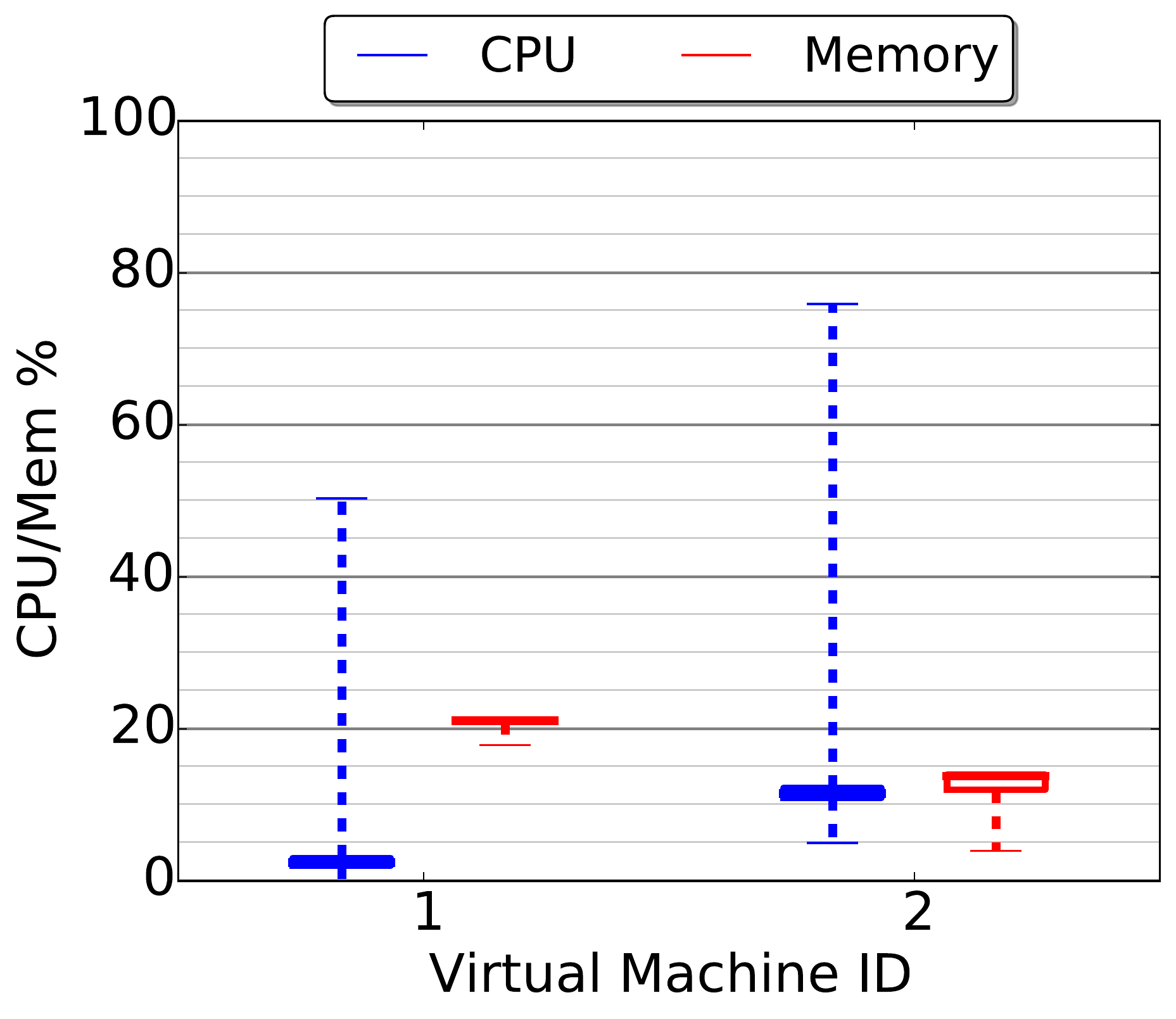}
		\label{fig:storm:train:city:cpu}
	}
	\subfloat[FIT]{
		\includegraphics[width=0.235\textwidth]{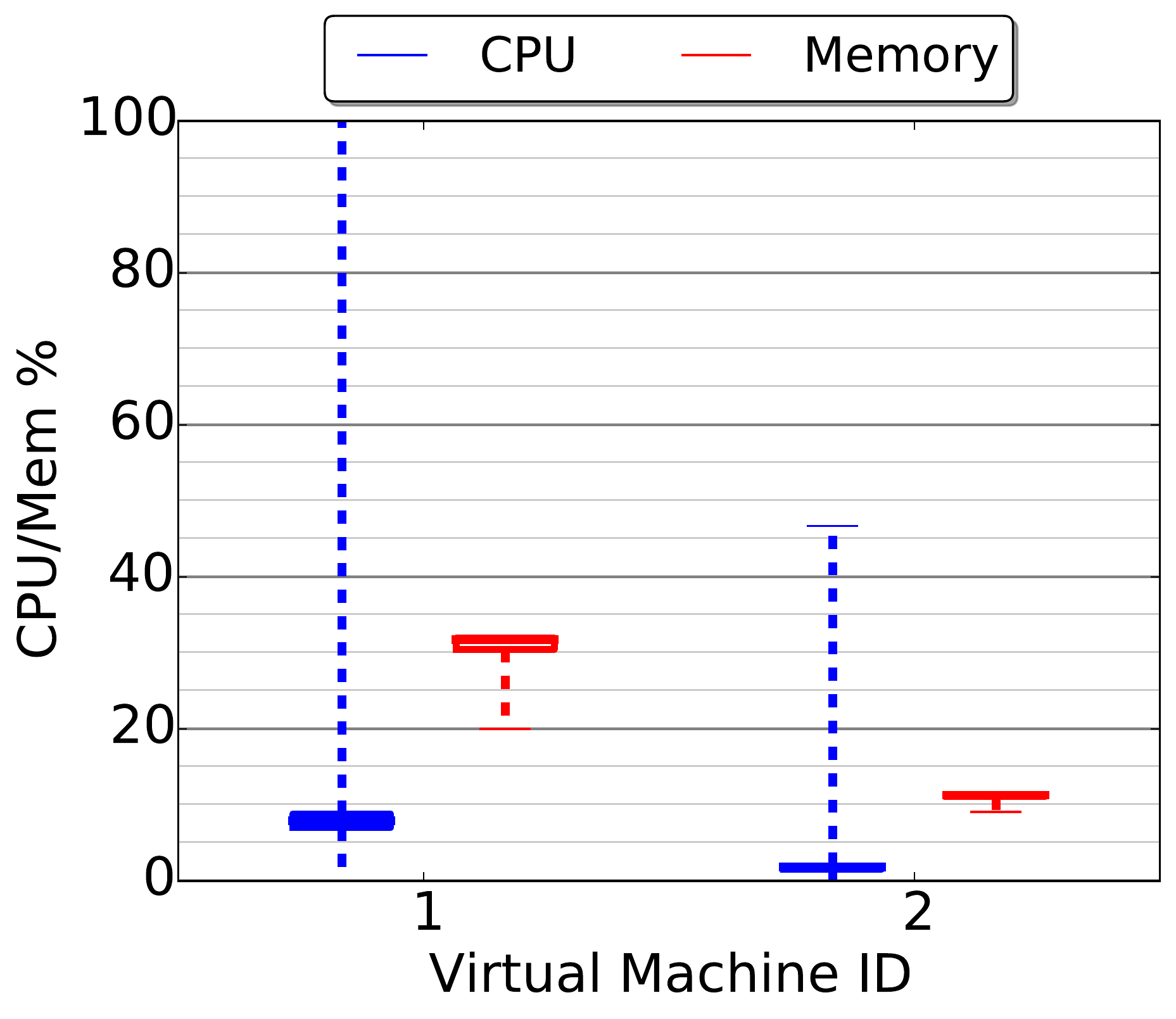}
		\label{fig:storm:train:fit:cpu}
	}
	\subfloat[TAXI]{
		\includegraphics[width=0.235\textwidth]{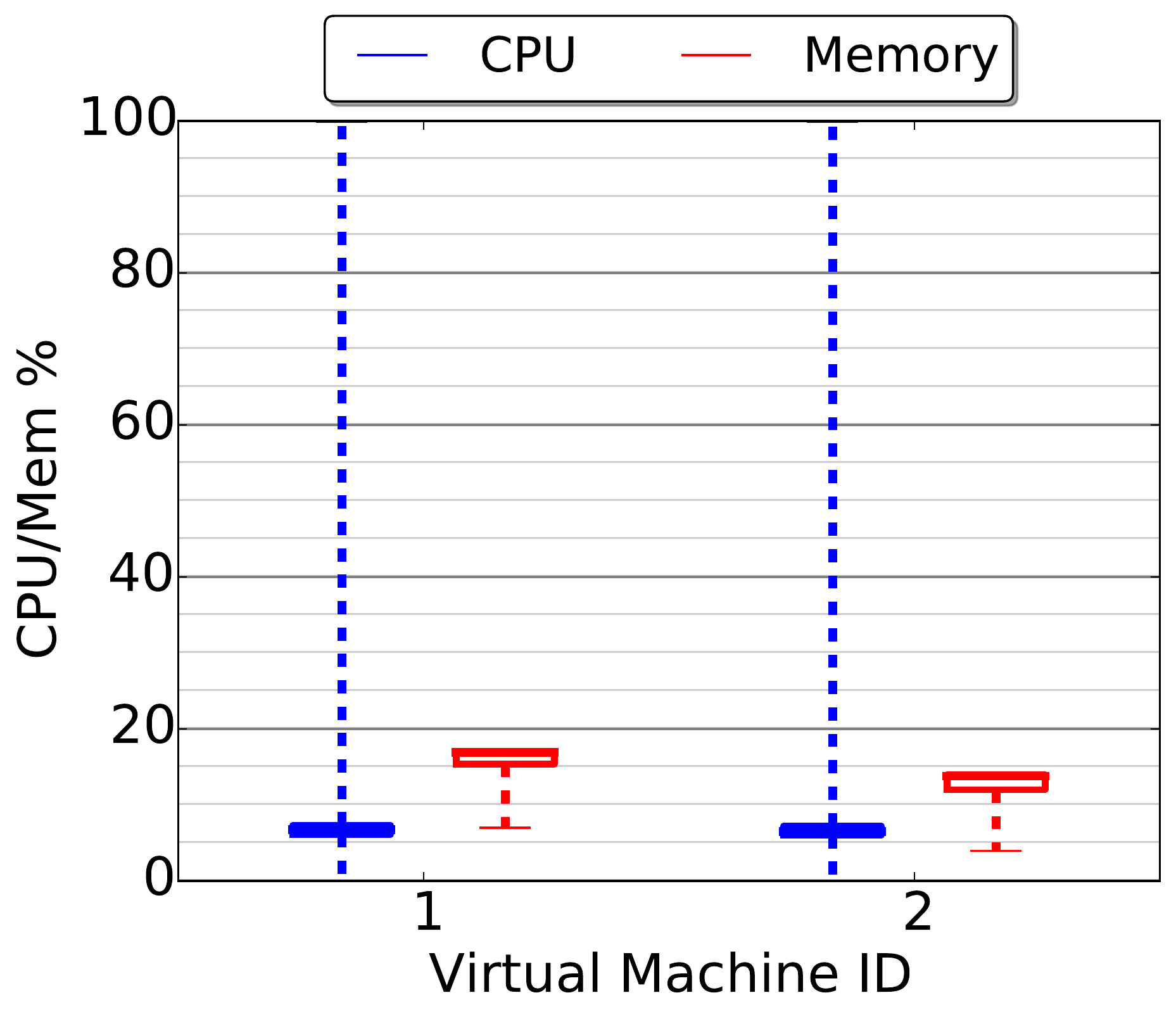}
		\label{fig:storm:train:taxi:cpu}
	}
	\caption{CPU and Memory utilization plots for \emph{TRAIN} application benchmarks three workloads, CITY FIT and TAXI. GRID workload is not used as it has only the target field, and no additional field to predict upon.}
	\label{fig:storm:train:cpu}
\end{figure}

\begin{figure}[t]
	\centering
	\subfloat[CITY]{
		\includegraphics[width=0.235\textwidth]{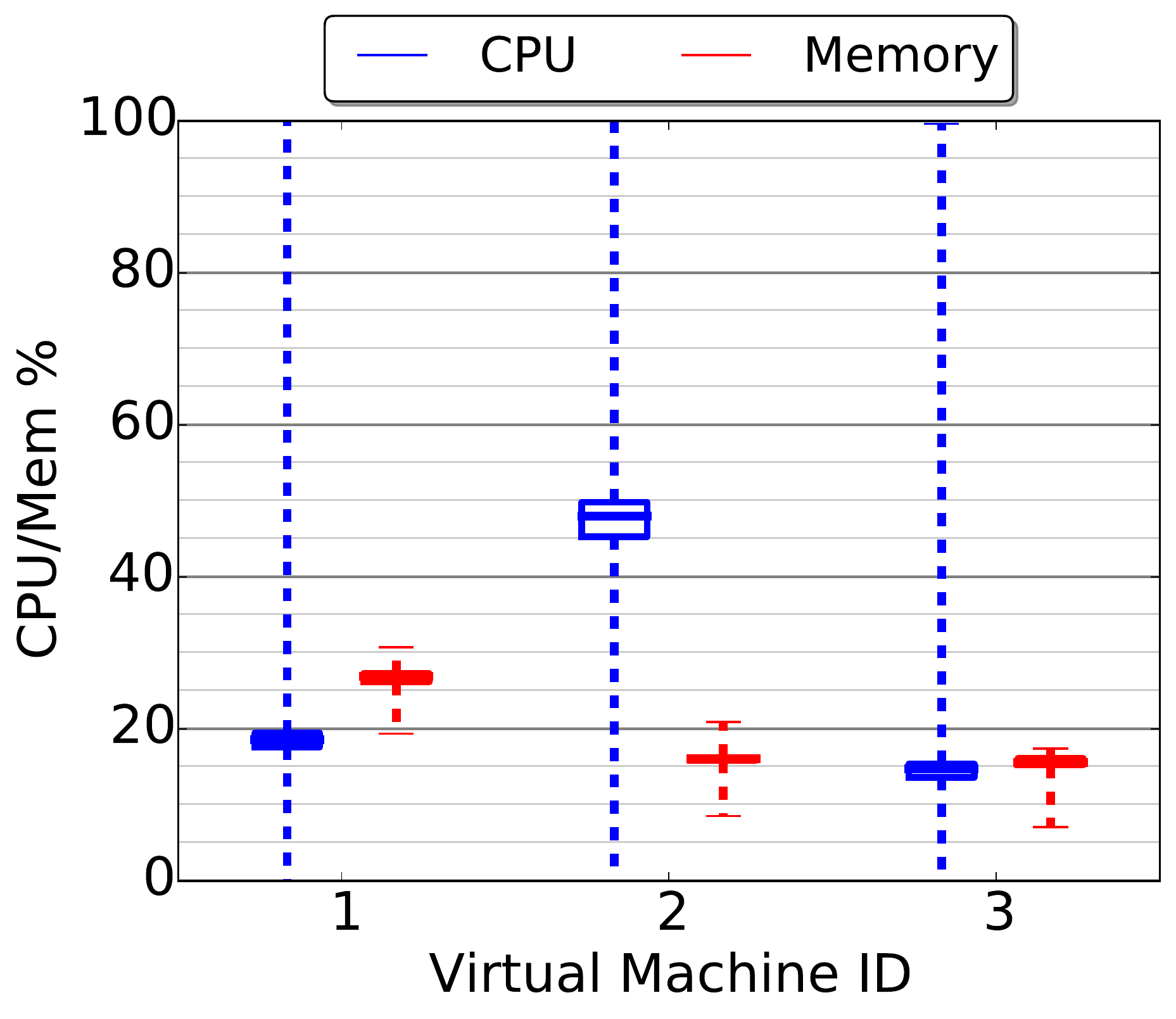}
		\label{fig:storm:pred:city:cpu}
	}
	\subfloat[FIT]{
		\includegraphics[width=0.235\textwidth]{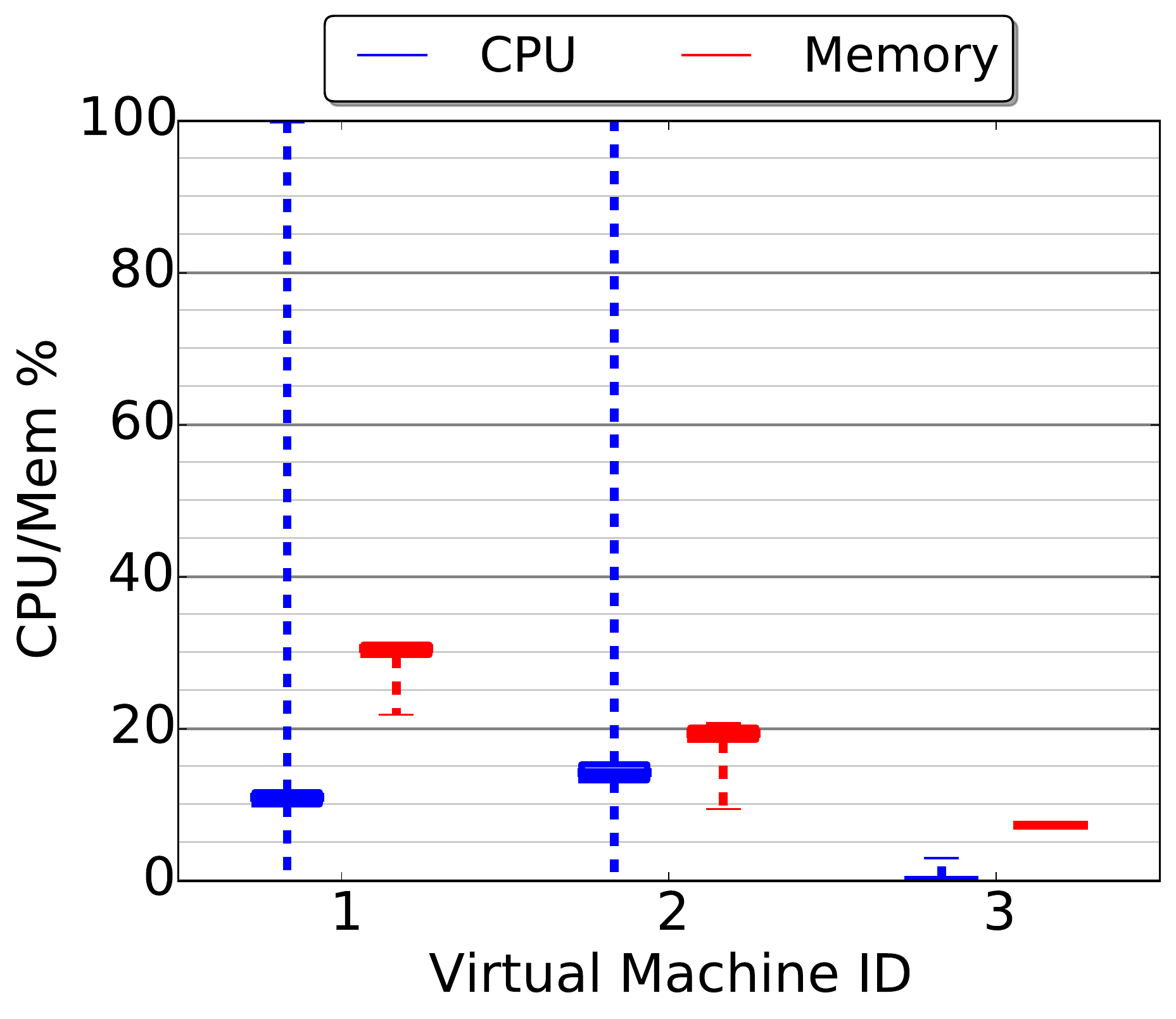}
		\label{fig:storm:pred:fit:cpu}
	}
	\subfloat[TAXI]{
		\includegraphics[width=0.235\textwidth]{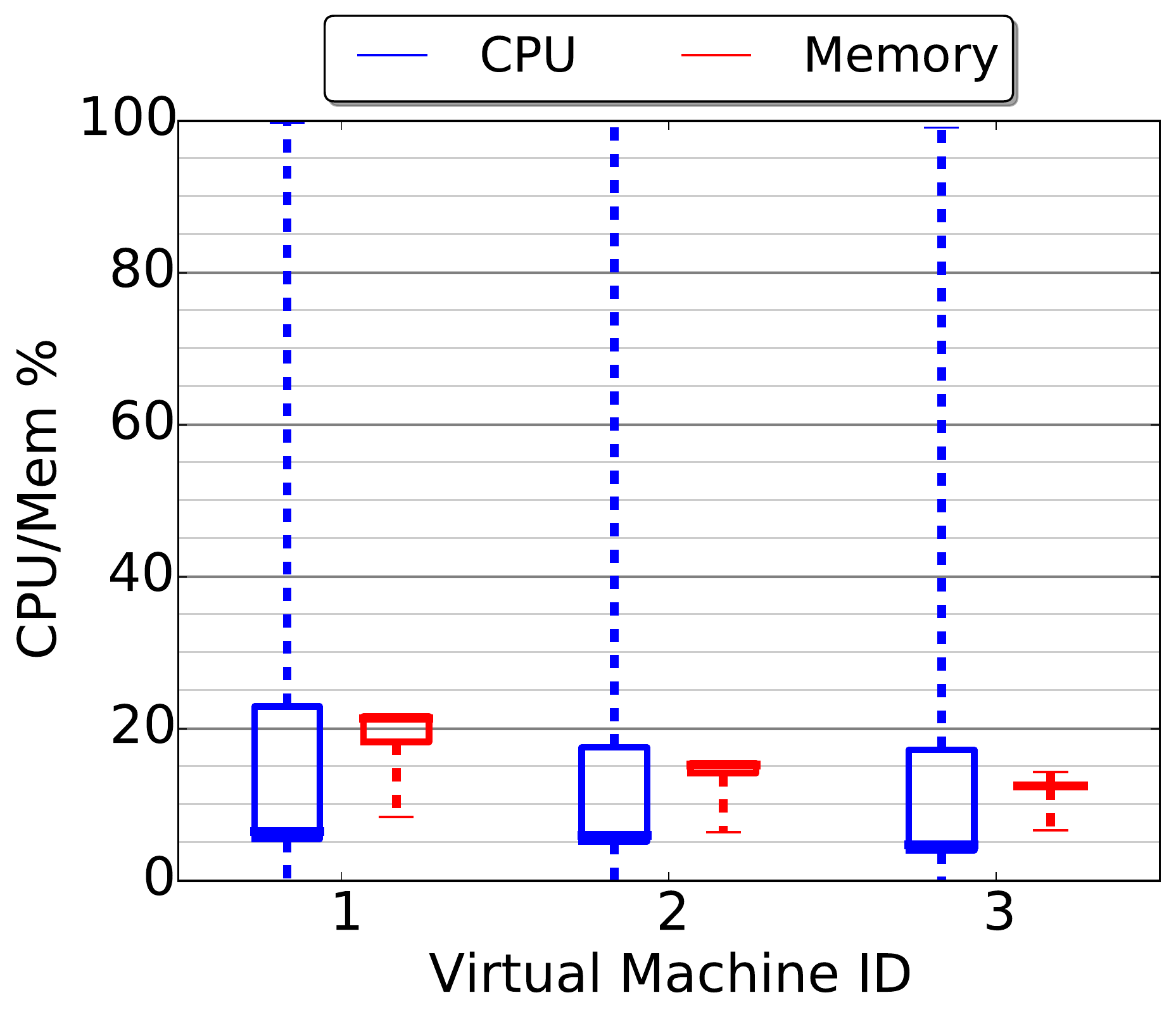}
		\label{fig:storm:pred:taxi:cpu}
	}
	\caption{CPU and Memory utilization plots for \emph{PRED} application benchmarks on three workloads, CITY FIT and TAXI. GRID workload is not used as it has only the target field, and no additional field to predict upon.}
	\label{fig:storm:pred:cpu}
\end{figure}


The ETL and STATS application benchmarks are run for the CITY, FIT, GRID and TAXI stream workloads. TRAIN and PRED are run for CITY, FIT and TAXI datasets and not for GRID because it has only one observation field, and prediction tasks such Decision Tree and Multivariate Linear Classifier uses a combination of fields to predict or classify an observational field. The input rate is as per scaling discussed in table ~\ref{tbl:datasets} for each dataset. 

The \emph{end-to-end latencies} of the applications depend on the sum of the end-to-end latencies of each task in the critical path of the dataflow.
For the ETL application, latency values in Fig.~\ref{fig:storm:etl:latency} remain the same $30~millisec$ for CITY, FIT and TAXI datasets. GRID has a higher variation in latency than others because of its normal distribution of messages over timestamp. The median latency for all the datasets are nearly comparable, with GRID having median latency $50~millisec$ and CITY, TAXI and FIT around $30~millisec$.
The STATS dataflow has latency values  in the range of $10-40~millisec$, shown in Fig.~\ref{fig:storm:stats:latency}, which is higher than ETL and PRED. This is mainly due to the GroupViZ meta-task which batches messages, forming a time-series for plotting, and then accumulating the plots to create a zipped file. Also, its median latency values are highly variable depending on the dataset. The reason is that the accumulation and plotting are done separately for every distinct sensor ID until a fixed count is reached, and hence the latency for the meta-task depends on the content of input messages received.

The TRAIN dataflow's timer source task simulates the model training trigger every $2~hours$ of original time for CITY, every day for TAXI, and every minute for the FIT dataset. This translates to a benchmark time period of $2-5$~mins between two source events. 
The latency values for TRAIN are understandably higher than other applications since it is a batch processing dataflow encoded as a streaming dataflow. The key reason is the Azure Table Range task that scans the full table to fetch rows that were inserted since the last training time. Also, the latency for the CITY dataset in Fig.~\ref{fig:storm:train:latency} is larger at $300~sec$ than FIT and TAXI datasets that are at $50~sec$ due to the difference in the table sizes. CITY has $3,629,428$ rows in its table while TAXI has $753,382$ rows inserted. 
The PRED topology's latencies (Fig.~\ref{fig:storm:pred:latency}) also remain close together at $20~millisec$ for all the datasets. The large range of whiskers for all datasets in PRED is due to DTC and MLR tasks, which exhibit significant variations in their runtimes even for the micro-benchmarks.

The \emph{jitter} is also close to zero in all cases (Fig.~\ref{fig:storm:app:jitter}), indicating a sustainable performance for the application benchmarks. The whiskers for STATS are not visible as the total number of messages at the sink tasks are comparatively fewer than the input messages since the 
GroupViZ task accumulates many of the inputs in singleton outputs per sensor ID.
Similarly, the whiskers for TRAIN are larger as few messages are emitted from source (max $10~msg/sec$ for FIT) in total, and thus most of the time variation is observed between source and sink rate.

The number of \emph{cores and VMs} required for the same application varies with the workload used (Tbl.~\ref{table:slots}). This is due to the difference in input rate that is processed by tasks for the respective workload, thus requiring different number of cores per task. We also see that the resource allocation strategy is generally liberal, and resources are under-utilized.
The CPU utilization for STATS is higher at $20-80\%$ than other applications (Fig.~\ref{fig:storm:stats:cpu}). This is due to AVG, DAC and GroupViZ tasks requiring higher CPU\%, matches with the CPU\% required for microbenchmarks. Also memory usage is higher for STATS in comparison to others due to GroupViZ task accumulating the messages 
and plots in memory before zipping (Fig.~\ref{fig:storm:stats:cpu}).
The CPU utilisation for TRAIN is fairly small due to the low message rate, and the memory usage is comparatively high at $20\%$ as the large batch of table rows is stored in memory for model training.
The CPU utilization for the FIT workload is the least for all the application benchmarks due to the fact that it has the least rate at $500~msg/sec$, and we have assigned exclusive an core to each of its tasks. TAXI has a low CPU usage, mostly at a $5\%$ median, with a wide box (Figs.~\ref{fig:storm:etl:taxi:cpu},~\ref{fig:storm:pred:taxi:cpu}~\ref{fig:storm:stats:taxi:cpu}~\ref{fig:storm:train:taxi:cpu}) -- this is due to its bi-modal distribution with low input rates at nights, with lower utilization, and high in the day with higher utilization.
In general, we see that such a resource under-utilization strongly motivates the need for robust resource allocation strategies for IoT applications on DSPS.

\section{Conclusion}
\label{sec:conclusion}
In this paper, we have proposed \emph{RIoTBench}, a novel benchmark suite for evaluating distributed stream processing systems for Internet of Things applications, which encompasses several emerging domains. Fast data platforms like DSPS are integral for the rapid decision making needs of IoT applications. Our proposed micro and application benchmarks help evaluate their efficacy using common tasks found in IoT domains, as well as fully-functional dataflows for pre-processing, statistical summarization and predictive analytics. These applications naturally fit into the OODA interaction model found in many IoT domains. These benchmarks are combined with four real-world data streams from Smart Grid, Smart Transportation, Urban Sensing and personal fitness domains of IoT, that are further spatially and temporally scaled to recreate the stream profiles of contemporary IoT deployments. The proposed benchmark has been validated for the highly-popular Apache Storm DSPS, and the performance metrics reported.

As future work, we would like to add event pattern detection and notifications as tasks to our benchmark suite to complete the representative categories. The benchmark can also be used to evaluate other popular DSPS such as Apache Spark Streaming and Flink. Incidentally, these tasks and applications we have provided have real and accurate business logic. Thus, they form a valuable library of tasks that can be used in both generic and IoT streaming applications. We are currently in the process of integrating customized versions of these benchmark applications into the IISc Smart Campus IoT project for smart water and power management~\footnote{IISc Smart Campus Project, \url{http://smartx.cds.iisc.ac.in}}.

\section*{Acknowledgments}
We acknowledge detailed inputs provided by Tarun Sharma of NVIDIA Corp. and formerly from IISc in preparing this paper. The experiments on Microsoft Azure were supported through a grant from Azure for Research. We thank the reviewers of the Technology Conference on Performance Evaluation \& Benchmarking (TPCTC), 2016, for their valuable comments to improve the benchmark suite.



\bibliographystyle{plain}
\footnotesize{
	\bibliography{main}
}

\clearpage
\appendix
\section{Configurations used in Application Dataflows}

\begin{table}[h]
	\centering
	\footnotesize
	\caption{Attributes Used in Tasks of the ETL Application}
	\label{tab:etl:field}
	\resizebox{\linewidth}{!}{%
	\begin{tabular}{c|p{2.95cm}p{2.95cm}p{2.95cm}p{3.95cm}}
		\hline
		\textbf{Task} &~\textbf{CITY} &~\textbf{FIT} &~\textbf{GRID} &~\textbf{TAXI}  \\ \hline \hline
		\textbf{ANN}$^*$  & location, sensor type  & age, gender & tariff allocation, sme allocation, stimulus allocation & driver name, city, company  \\\hline 
		\textbf{BLF}      & source  & N/A~$^\dagger$ & meterid  & taxi identifier  \\\hline 
		\textbf{INP}      & temperature, humidity, light, dust, airquality raw & acceleration chest, arm, ankle X/Y/Z, ecg 1 & energyConsumed & N/A~$^\ddagger$ \\ \hline                                                                  
		\textbf{RGF}      & temperature, humidity, light, dust, airquality raw & acceleration chest, arm, ankle X/Y/Z, ecg 1 & energyConsumed & trip time in sec, trip distance, fare amount, surcharge, mta tax, tip amount, tolls amount, total amount \\ \hline
		\hline
		\multicolumn{5}{l}{\emph{$^*$ Annotation attributes that are added to the dataset by ANN, either provided with the dataset or synthetically}}\\
		\multicolumn{5}{l}{\emph{$^\dagger$ No fields were used for the particular task with the dataset because the number of unique subjects is very less (10) for FIT thus not requires BLF.}}\\
		\multicolumn{5}{l}{\emph{$^\ddagger$ Interpolation of values over different Taxi trips is not meaningful.}}\\
	\end{tabular}}
\end{table}

\begin{table}[h]
	\centering
	\footnotesize
	\caption{Attributes Used in Tasks of the STATS Application}
	\label{tab:stats:field}
	\resizebox{\linewidth}{!}{%
	\begin{tabular}{c|p{2.95cm}p{2.95cm}p{2.95cm}p{3.95cm}}
		\hline
		\textbf{Task} &~\textbf{CITY} &~\textbf{FIT} &~\textbf{GRID} &~\textbf{TAXI}  \\ \hline \hline                                                                                                                                 
		\textbf{AVG} &temperature, humidity, light, dust, airquality raw  & acceleration chest, arm, ankle X/Y/Z, ecg 1/2 & energyConsumed & trip time in sec, trip distance, fare amount, surcharge, mta tax, tip amount, tolls amount, total amount\\\hline
		\textbf{DAC} & temperature & ecg 1 & energyConsumed & N/A~$^\dagger$ \\ \hline
		\textbf{SLR}&temperature, humidity, light, dust, airquality raw & acceleration chest, arm, ankle X/Y/Z, ecg 1/2 & energyConsumed & trip time in sec, trip distance, fare amount, surcharge, mta tax, tip amount, tolls amount, total amount \\ \hline
		\hline
		\multicolumn{5}{l}{\emph{$^\dagger$ No fields were used for the particular task with the dataset because DAC over individual Taxi trips is not meaningful.}}
	\end{tabular}}
\end{table}
\begin{table}[h]
	\centering
	\footnotesize
	\caption{Attributes Used in Tasks of the PRED Application}
	\label{tab:pred:field}
	\resizebox{\linewidth}{!}{%
	\begin{tabular}{c|p{2.95cm}p{2.95cm}p{2.95cm}p{3.95cm}}
		\hline
		\textbf{Task} &~\textbf{CITY} &~\textbf{FIT} &~\textbf{GRID} &~\textbf{TAXI}  \\ \hline \hline
		\textbf{AVG} & airquality raw & ecg 1 & N/A~$^\dagger$  & fare amount \\ \hline 
		\textbf{DTC} & $\mathcal{F}$(temperature, humidity, light, dust, airquality raw) $\rightarrow \{C1 \mid C2 \mid C3 \mid C4 \}^*$ & $\mathcal{F}$(acceleration chest, arm, ankle X/Y/Z, ecg 1) $\rightarrow \{C1 \mid C2 \mid C3 \mid C4 \}^*$ & N/A~$^\dagger$   & $\mathcal{F}$(trip time in sec, trip distance, fare amount) $\rightarrow \{C1 \mid C2 \mid C3 \mid C4 \}^*$ \\ \hline
		\textbf{MLR} & $\mathcal{F}$(temperature,  humidity,  light)  $\rightarrow$  airquality raw & (acceleration chest, arm, ankle X/Y/Z) $\rightarrow$ ecg 1 & N/A~$^\dagger$ & $\mathcal{F}$(trip time in secs, trip distance)$\rightarrow$ fare amount  \\ \hline
		\hline
		\multicolumn{5}{l}{\emph{$^*$ Classses used for prediction by DTC task}}\\
		\multicolumn{5}{l}{\emph{$^\dagger$ No fields were used for the particular task with the dataset as GRID is univariate whereas DTC and MLR tasks require multiple fields.}}
	\end{tabular}
}
\end{table}
\begin{table}[h]
	\centering
	\footnotesize
	\caption{Attributes Used in Tasks of the TRAIN Application}
	\label{tab:train:field}
	\resizebox{\linewidth}{!}{%
	\begin{tabular}{c|p{2.95cm}p{2.95cm}p{2.95cm}p{3.95cm}}
		\hline
		\textbf{Task} &~\textbf{CITY} &~\textbf{FIT} &~\textbf{GRID} &~\textbf{TAXI}  \\ \hline \hline
		\textbf{DTT} & temperature, humidity, light, dust, airquality raw & acceleration chest, arm, ankle X/Y/Z, ecg 1  & N/A~$^\dagger$   & trip time in sec, trip distance, fare amount \\ \hline
		\textbf{MLT} & temperature, humidity, light, dust, airquality raw & acceleration chest, arm, ankle X/Y/Z, ecg 1 & N/A~$^\dagger$  & trip time in sec, trip distance,fare amount \\ \hline
		\hline
		\multicolumn{5}{l}{\emph{$^\dagger$ No fields were used for the particular task with the dataset as GRID is univariate whereas DTT and MLT tasks require multiple fields.}}
	\end{tabular}}
\end{table}

\end{document}